\documentclass[final,12pt, authoryear]{elsarticle}

\usepackage{url}
\usepackage[margin=0.75in]{geometry} 

\usepackage{float}
\usepackage[hidelinks,colorlinks=true,linkcolor=blue,citecolor=blue]{hyperref}
\usepackage{amsmath}
\usepackage[graphicx]{realboxes}
\usepackage{adjustbox}
\usepackage[table,xcdraw]{xcolor}
\usepackage[T1]{fontenc}
\usepackage{epsf}
\usepackage{caption}
\usepackage{subcaption}
\usepackage{booktabs}

\newcommand{\probP}{\text{I\kern-0.15em P}}
\usepackage{pifont}

\usepackage{amssymb}

\journal{}

\begin{document}

\begin{frontmatter}



\title{Deep Learning Meets Queue-Reactive: A Framework for Realistic Limit Order Book Simulation}



\author[inst1,inst2]{Hamza Bodor}
\ead{bodor.hamza@gmail.com}

\author[inst2]{Laurent Carlier}
\ead{laurent.carlier@bnpparibas.com}

\affiliation[inst1]{organization={Université Paris 1 Panthéon-Sorbonne, Centre d’Economie de la Sorbonne},
            addressline={106 Boulevard de l'Hôpital}, 
            city={Paris Cedex 13},
            postcode={75642},
            country={France}}

\affiliation[inst2]{organization={BNP Paribas Corporate and Institutional Banking, Global Markets Data \& Artificial Intelligence Lab},
            addressline={20 boulevard des Italiens}, 
            city={Paris},
            postcode={75009},
            country={France}}

\begin{abstract}

The Queue-Reactive model introduced by~\cite{huang2015simulating} has become a standard tool for limit order book modeling, widely adopted by both researchers and practitioners for its simplicity and effectiveness. We present the Multidimensional Deep Queue-Reactive (MDQR) model, which extends this framework in three ways: it relaxes the assumption of queue independence, enriches the state space with market features, and models the distribution of order sizes. Through a neural network architecture, the model learns complex dependencies between different price levels and adapts to varying market conditions, while preserving the interpretable point-process foundation of the original framework. Using data from the Bund futures market, we show that MDQR captures key market properties including the square-root law of market impact, cross-queue correlations, and realistic order size patterns. The model demonstrates particular strength in reproducing both conditional and stationary distributions of order sizes, as well as various stylized facts of market microstructure. The model achieves this while maintaining the computational efficiency needed for practical applications such as strategy development through reinforcement learning or realistic backtesting.

\end{abstract}




\end{frontmatter}


\newpage

\section{Introduction and Motivation}

Financial markets have evolved significantly over the past decades, with electronic limit order books (LOBs) becoming the predominant mechanism for price formation and trading across major exchanges worldwide. \citet{gould2013limit} provide a comprehensive overview of these markets, where participants interact by submitting limit orders (indicating willingness to trade at specific prices) and market orders (executing against existing limit orders), creating a complex and high-dimensional system whose dynamics emerge from the interactions of diverse market participants \citep{cont2011statistical}.

Understanding and modeling these dynamics is crucial for several reasons. First, market participants need realistic simulators to develop and test trading strategies before deploying them in live markets, where testing can be both risky and costly \citep{cartea2015algorithmic}. Second, regulators require tools to assess market stability and evaluate potential policy changes, as demonstrated by \citet{kirilenko2017flash} in their analysis of the Flash Crash. Third, researchers seek to understand the fundamental mechanisms driving price formation and market impact \citep{gatheral2010no}.

The complexity of limit order markets poses significant modeling challenges. Order arrivals exhibit complex temporal dependencies, with order flow at one price level potentially influencing dynamics at other levels. Order sizes vary considerably and show distinctive patterns. Furthermore, the market's response to large trades manifests through intricate feedback mechanisms, resulting in well-documented phenomena such as the ``square-root law'' of market impact \citep{gatheral2010no, cont2014price}.

Traditional approaches to modeling LOBs often rely on simplifying assumptions to maintain tractability. \citet{huang2015simulating} made a significant advance with their Queue-Reactive (QR) model by capturing the dependency between order flow and queue sizes while maintaining analytical tractability. Recently, \citet{bodor2024novel} extensively investigated the performance of the QR model against multiple stylized facts of the Bund Market and proposed several extensions, particularly focusing on order size modeling. However, these enhanced models still maintain strong independence assumptions between price levels and simplified order size distributions.

The emergence of deep learning techniques offers new possibilities for capturing these complex dependencies while maintaining computational feasibility. Recent successes in applying neural networks to financial markets, as demonstrated by \citet{sirignano2019universal}, suggest that these methods could help overcome the limitations of traditional approaches while preserving their interpretability.

In this paper, we propose a deep learning extension of the Queue-Reactive model that significantly increases its flexibility while maintaining its economic interpretability. Our approach combines the theoretical insights of the QR framework with the representational power of deep neural networks. The key contributions of our work include:
\begin{enumerate}
    \item A neural network extension of the Queue-Reactive framework for order book modeling balancing interpretability and computational efficiency;
    \item Integration of cross-price level dependencies and realistic order size modeling;
    \item Empirical validation demonstrating the model's ability to reproduce sophisticated market impact patterns and key stylized facts;
    \item A framework that bridges the gap between traditional stochastic models and modern machine learning approaches.
\end{enumerate}

The remainder of this paper is organized as follows. Section 2 reviews the literature on limit order book modeling. Section 3 presents the Queue-Reactive model, its extensions, and limitations. Section 4 introduces the MDQR framework and analyzes its properties. Section 5 compares our approach with related work. Section 6 concludes with directions for future research.

\section{Related Work}

The modeling of limit order book dynamics has attracted significant attention from both academics and practitioners, leading to diverse approaches ranging from purely statistical models to sophisticated machine learning techniques. Early empirical studies by \citet{bouchaud2002statistical} and \citet{potters2003more} established key stylized facts about order flow, price formation, and market impact that subsequent models sought to reproduce.

Statistical approaches, particularly those based on point processes, formed the initial framework for modeling order book dynamics. \citet{smith2003statistical} and \citet{cont2010stochastic} introduced simple Poisson process models that provided valuable analytical insights while maintaining tractability, though they failed to capture the complex dependencies observed in empirical data. This limitation led to the development of more sophisticated models using Hawkes processes, as demonstrated by \citet{abergel2013mathematical, abergel2015long}, which better captured the clustering of events and temporal dependencies in order flow. A significant advancement came when \citet{huang2015simulating} introduced the Queue-Reactive model, making arrival intensities dependent on queue sizes and better reflecting the empirical relationship between order flow and market state. Recently, \citet{bodor2024novel} extended this framework by incorporating order sizes in the model, demonstrating improved empirical fit.

\citet{paddrik2012agent} offered a different perspective through agent-based models focusing on the behavior of market participants. While these models provided valuable insights into market mechanisms, particularly in stress scenarios like the Flash Crash studied by \citet{kirilenko2017flash}, the challenge of calibrating multiple agent behaviors to match empirical data has limited their practical application \citep{vyetrenko2020get}. Moreover, as highlighted by \citet{axtell2022agent}, the evolution of agent-based modeling in economics and finance underscores its potential to capture complex interactions and emergent phenomena, although practical challenges in implementation persist.

The emergence of machine learning techniques has opened new avenues for LOB modeling. \citet{sirignano2019universal} demonstrated the power of deep learning in forecasting price movements, while recent work has focused on generative modeling of entire order book dynamics. Notably, \citet{coletta2022conditional} developed a GAN-based approach for LOB simulation, focusing on explainability and robustness. Their work provided frameworks for evaluating generator quality and ensuring stable training, critical considerations for practical applications. Building on this foundation, \citet{cont2023simulation} extended the GAN-based approach, successfully reproducing complex market behaviors, including the square-root law of market impact. A particularly significant contribution came from \citet{hultin2023generative}, who adopted a recurrent neural network-based approach that successfully captured temporal dependencies while preserving key statistical properties of order book dynamics.

Recent work has also explored hybrid approaches that combine traditional modeling frameworks with machine learning techniques. \citet{cont2021stochastic} modeled the LOB as a density evolving according to a stochastic partial differential equation, while using neural networks to capture complex dependencies in the driving processes. These hybrid approaches suggest the potential for combining the interpretability of traditional models with the flexibility of modern machine learning methods.

Our work builds on these developments by proposing a deep learning extension of the Queue-Reactive model that maintains its economic interpretability while significantly increasing its flexibility to capture complex market dynamics. This approach bridges the gap between traditional stochastic models and modern machine learning techniques, offering a framework that benefits from both traditions.

\section{The Queue-Reactive Model and Its Extensions}


The Queue-Reactive (QR) model, introduced by \citet{huang2015simulating}, provides an elegant framework for modeling limit order book dynamics. At its core, the model views the limit order book as a collection of queues centered around a reference price, typically the mid-price. Each queue represents the accumulated volume of orders at a specific price level, with the book divided into bid and ask sides. The evolution of these queues is driven by three types of events: limit orders that add to existing queues, cancellations that remove orders, and market orders that execute against available liquidity.

What distinguishes the QR model from earlier approaches is its central insight: the arrival rates of different order types depend on the current state of the queues. This state dependency captures a key empirical observation - market participants adjust their behavior based on visible order book conditions. For instance, a large queue at the best bid price might attract more market sell orders, while simultaneously discouraging additional limit buy orders at that level.

The model represents the limit order book as a $2K$-dimensional vector, where $K$ denotes the number of price levels tracked on each side of the book. The reference price \(p_{ref}\) serves as the center point, dividing the book into bid side queues \([Q_{-i} : i = 1, ..., K]\) and ask side queues \([Q_i : i = 1, ..., K]\). Here, \(Q_{\pm i}\) represents the queue size at a distance of \(i-0.5\) ticks from the reference price, with negative indices for the bid side and positive for the ask side.

The mathematical formulation of the model captures this intuition through a system of interacting point processes. Let us consider a sequence of events occurung at a given queue \(\mathcal{E} = \{e_k\}_{k=1}^N\), where each event \(e_k\) is characterized by its type \(\eta_k \in \{\mathrm{L}, \mathrm{C}, \mathrm{M}\}\) (limit, cancel, or market), the queue size \(q_k\) just before the event, and the time interval \(\Delta t_k\) since the previous event.

The key component of the model is the specification of intensity functions \(\lambda^\eta(q)\) for each event type \(\eta\), which determine the arrival rates of different orders based on the current queue size. These intensities combine to form a global intensity, which is the intensity of queue update (regardless of which event is updating):

\[
\Lambda(q) \;=\; \lambda^{\mathrm{L}}(q) \;+\; \lambda^{\mathrm{C}}(q) \;+\; \lambda^{\mathrm{M}}(q).
\]\\

The model's likelihood function then takes the form:

\[
\mathcal{L}(\{\lambda^\eta\} \mid \mathcal{E}) \;=\; \prod_{k=1}^{K} e^{-\Lambda(q_k)\,\Delta t_k} \; \lambda^{\eta_k}(q_k).
\]

The maximizer of this likelihood can be derived analytically and has the form:

\begin{equation}
\hat{\lambda}^\eta(n) = \frac{\# \{e_k \in \mathcal{E} \mid \eta_k = \eta, q_k = n\}}{\# \{e_k \in \mathcal{E} \mid q_k = n\}} \cdot \left( \frac{1}{\#\{k \mid q_k = n\}} \sum_{\{k \mid q_k = n\}} \Delta t_k \right)^{-1}
\end{equation}

This formulation has proven remarkably successful in reproducing several key features of limit order markets, including the relationship between queue sizes and order flow, and the mean-reversion behavior of queue sizes. However, the base model makes several simplifying assumptions that limit its ability to capture more complex market dynamics. Most notably, it assumes independence between price levels and constant order sizes within each level.

In \cite{bodor2024novel}, we extend the Queue-Reactive model by incorporating order sizes into the intensity functions, modifying them to take the form:
\begin{equation}
\hat{\lambda}^{\eta, s}(n) = \frac{\# \{e_k \in \mathcal{E} \mid \eta_k = \eta, s_k=s, q_k = n\}}{\# \{e_k \in \mathcal{E} \mid q_k = n\}} \cdot \left( \frac{1}{\#\{k \mid q_k = n\}} \sum_{\{k \mid q_k = n\}} \Delta t_k \right)^{-1}
\end{equation}

This extension, known as SAQR, achieves market simulations with volatility levels closer to historical data while preserving other important properties like the distribution of order sizes. However, both the original QR model and SAQR rely solely on queue sizes as features for order arrival modeling, limiting their ability to reproduce certain important stylized facts, such as excitation between events.

These limitations motivate our extension of the QR framework to accommodate a broader state space. By leveraging deep learning techniques, we can maintain the model's interpretable structure while incorporating more realistic features such as cross-price level dependencies and varying order sizes. 

\subsection{Deep Queue-Reactive Model}

While the QR model captures queue dynamics through its dependency on queue sizes with the likelihood function
\[
\mathcal{L}(\{\lambda^\eta\} \mid \mathcal{E}) \;=\; \prod_{k=1}^{N} e^{-\Lambda(q_k)\,\Delta t_k} \; \lambda^{\eta_k}(q_k),
\]
market participants' decisions depend on broader market conditions. The Deep Queue-Reactive (DQR) model generalizes this framework by replacing the queue-size dependency $q_k$ with a flexible state vector $x_k$ that can incorporate any relevant market features.

The model parameterizes the intensity functions through neural networks $\lambda_\theta^{\eta}(x_k)$, where $\theta$ represents the network parameters. Taking the logarithm of the likelihood:
\[
l(\lambda_\theta \mid \mathcal{E}) = \sum_{k=1}^N \log \lambda_\theta^{\eta_k}(x_k) - \Lambda_\theta(x_k)\Delta t_k
\]
where $\Lambda_\theta(x_k) = \sum_{\eta \in \{\mathrm{L,C,M}\}} \lambda_\theta^{\eta}(x_k)$.\\

The model is calibrated by minimizing the negative log-likelihood:
\[
\textrm{loss}(\theta \mid \mathcal{E}) = -l(\lambda_\theta \mid \mathcal{E}) = \sum_{k=1}^N \Lambda_\theta(x_k)\Delta t_k - \log \lambda_\theta^{\eta_k}(x_k)
\]

This neural network parameterization enables learning complex relationships between market conditions and order flow, while the flexibility in defining $x_k$ allows incorporation of any relevant market features beyond queue sizes. Section \ref{sec:DQR_results} demonstrates how different choices of state space variables affect the simulated market properties.

\subsection{Data}
\label{sec:Data1}

The calibration data is derived from Euro-Bund Futures (FGBL), specifically covering active trading days over three months from March to June 2022 for futures maturing in June 2022 (FGBLM2). The dataset is filtered to include periods from 9 AM to 6 PM, with a maximum depth of five levels on each side of the order book.\footnote{Euro-Bund Futures are derivative contracts tied to 10-year German government bonds (Bund) and traded on the Eurex Exchange. Each contract, featuring a notional coupon rate of 6\%, is designed to mature between 8.5 and 10.5 years on the delivery day.}



\begin{table}[H]
\centering
\begin{tabular}{llllllll}
\cline{1-6}
\multicolumn{1}{l}{\textbf{Level}} & \multicolumn{1}{c}{\textbf{\begin{tabular}[c]{@{}c@{}}\#L\\ $(\times 10 ^6)$\end{tabular}}} & \multicolumn{1}{c}{\textbf{\begin{tabular}[c]{@{}c@{}}\#C\\ $(\times 10 ^6)$\end{tabular}}} & \multicolumn{1}{c}{\textbf{\begin{tabular}[c]{@{}c@{}}\#M\\ $(\times 10 ^4)$\end{tabular}}} & \multicolumn{1}{c}{\textbf{AES}} & \multicolumn{1}{c}{\begin{tabular}[c]{@{}c@{}}\textbf{AIT}\\ (ms)\end{tabular}} & \multicolumn{1}{c}{} & \multicolumn{1}{c}{} \\ \cline{1-6}
1                    & 32.9                                                                             & 30.1                                                                             & 212                                                                              & 6.25                             & 57.3                                                                              &                      &                      \\ \cline{1-6}
2                    & 6.62                                                                             & 5.63                                                                            & 0.51                                                                             & 5.04                             & 293                                                                             &                      &                      \\ \cline{1-6}
3                    & 4.59                                                                             & 4.58                                                                           & 0                                                                                & 6.16                          & 392                                                                             &                      &                      \\ \cline{1-6}
4                    & 2.79                                                                            & 2.83                                                                           & 0                                                                                & 5.69                             & 537                                                                             &                      &                      \\ \cline{1-6}
5                    & 3.93                                                                           & 3.63                                                                           & 0                                                                                & 3.39                           & 443                                                                             &                      &                      \\ \cline{1-6}
\end{tabular}
\caption{Descriptive statistics about events for each level.}
\label{tab:stat_desc}
\end{table}

Statistical characteristics of the dataset used for model calibration are summarized in Table~\ref{tab:stat_desc}, including event counts by type, Average Event Size (AES), and Average Inter-event Time (AIT) for each queue. The data reveals a concentration of activity near the mid-price, particularly at the best bid and ask levels.

Our dataset combines limit order book updates with daily trading records. The calibration process requires precise order flow information, which we extract by classifying each update as either liquidity provision (limit orders) or consumption (cancellations and market orders), following the methodology detailed in~\cite{bodor2024stylized}. A custom matching engine processes orders generated by each model to maintain a consistent simulated order book state.

Following~\cite{huang2015simulating}, we apply several preprocessing steps:
\begin{itemize}
\item Queue sizes are normalized by their respective average event sizes: for queue $i$ with size $q_i$ and average event size $AES_i$, we compute $q_i \leftarrow \lceil \frac{q_i}{AES_i} \rceil$.
\item We partition the data into segments of constant reference price, resetting the inter-event time measurements across all queues when the reference price changes.
\end{itemize}

Several important aspects characterize our simulation approach:
\begin{itemize}
\item Event timing and type determination using calibrated intensity functions
\item Price evolution following the specified methodology, while maintaining endogenous market dynamics by avoiding artificial order book redraws
\item Queue size initialization for new price levels based on their empirical distributions after reference price changes
\end{itemize}

\subsection{Neural Architecture and Training Methodology}
\label{sec:DQR_architecture}
The model employs a neural network architecture to calibrate parametric intensities across price levels, balancing predictive performance with computational requirements. The architecture consists of a Multi-Layer Perceptron (MLP) with two hidden layers of dimensions 128 and 32 neurons using hyperbolic tangent (\textit{tanh}) activations, followed by an output layer of dimension 3 with Rectified Linear Unit (ReLU) activation that produces arrival intensities for limit orders, cancellations, and market orders. This configuration emerged from empirical testing, offering an effective balance between model expressiveness and computational efficiency-an important consideration for the simulation framework.

To enhance the model's learning capabilities and stability, we incorporate batch normalization layers between successive dense layers. In the cases when the input vector $x_k$ contains categorical features, we implement learnable embedding layers of dimension~2, allowing the model to discover meaningful representations of discrete variables while maintaining a compact parameter space.\\

The training protocol follows the subsequent methodology:

\begin{itemize}
    \item Data partitioning: 80\% of the dataset is allocated for training, with the remaining 20\% reserved for validation and early stopping criteria
    \item Optimization: Implementation of the Adam optimizer with cyclic learning rate scheduling, oscillating between $10^{-5}$ and $10^{-3}$
    \item Early stopping: Training termination after 10 consecutive epochs without improvement in validation loss
\end{itemize}

Table~\ref{tab:architecture} summarizes the key architectural components and training parameters:

\begin{table}[h]
\centering
\caption{Neural Network Architecture and Training Specifications}
\label{tab:architecture}
\begin{tabular}{ll}
\hline
\textbf{Component} & \textbf{Specification} \\
\hline
Hidden Layers & [128, 32] \\
Output Dimension & 3 (intensity values) \\
Embedding Dimension & 2 (categorical features) \\
Optimizer & Adam with cyclic LR \\
Learning Rate Range & $10^{-5}$ to $10^{-3}$ \\
\hline
\end{tabular}
\end{table}

This architecture demonstrates robust convergence properties while maintaining the computational efficiency necessary for practical deployment in high-frequency market scenarios.

\subsection{Examples of Applications}
\label{sec:DQR_results}

\subsubsection{Excitation between Events}
In financial markets, the type of the previous event ($\eta_{k-1}$) often influences the type and intensity of the next event. This phenomenon, known as excitation, is critical for realistic modeling of event dynamics. Traditional Queue-Reactive (QR) and Size-Aware Queue-Reactive (SAQR) models fail to capture excitation because they do not explicitly model any information about the order flow history, resulting in inaccurate excitation values.

To address this, we incorporated \( \eta_{k-1} \) as a feature in the state space \( x_k = [q_k, \eta_{k-1}] \) for the Deep Queue-Reactive (DQR) model. This inclusion enables the model to learn how the type of the last event influences subsequent events.

\begin{figure}[H]
    \centering
    \begin{subfigure}[b]{0.32\textwidth}
        \includegraphics[width=\textwidth,clip]{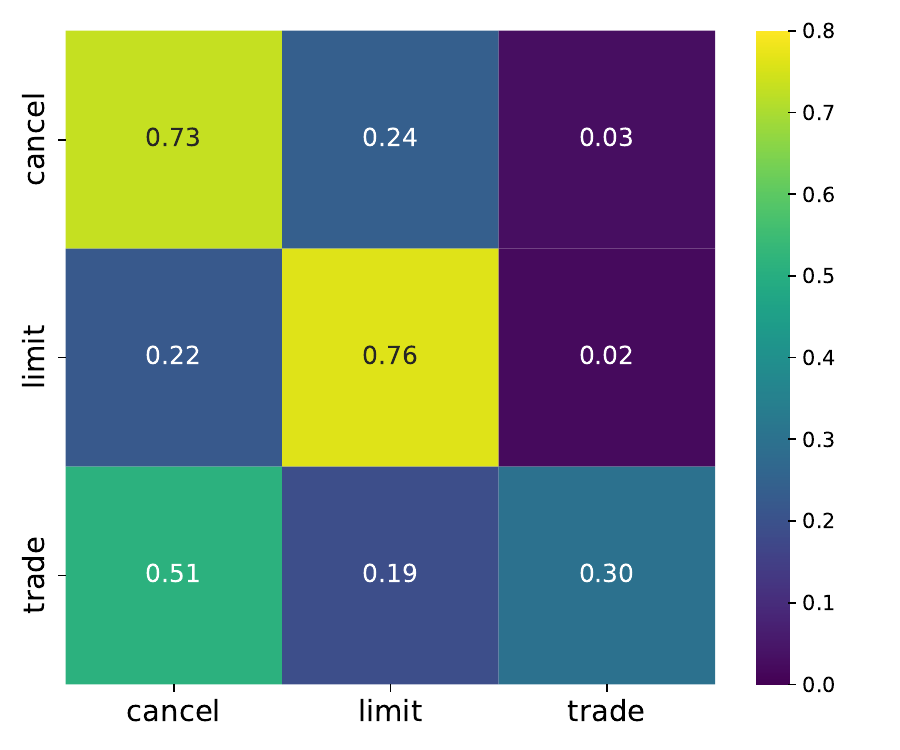}
        \caption{Real}
        \label{fig:sub1}
    \end{subfigure}
    \hfill
    \begin{subfigure}[b]{0.32\textwidth}
        \includegraphics[width=\textwidth,clip]{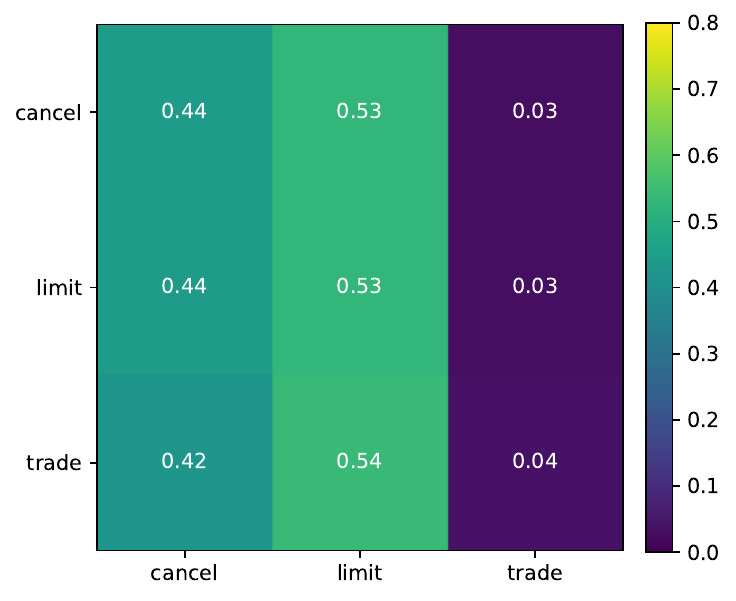}
        \caption{SAQR \& QR}
        \label{fig:sub2}
    \end{subfigure}
    \hfill
    \begin{subfigure}[b]{0.32\textwidth}
        \includegraphics[width=\textwidth,clip]{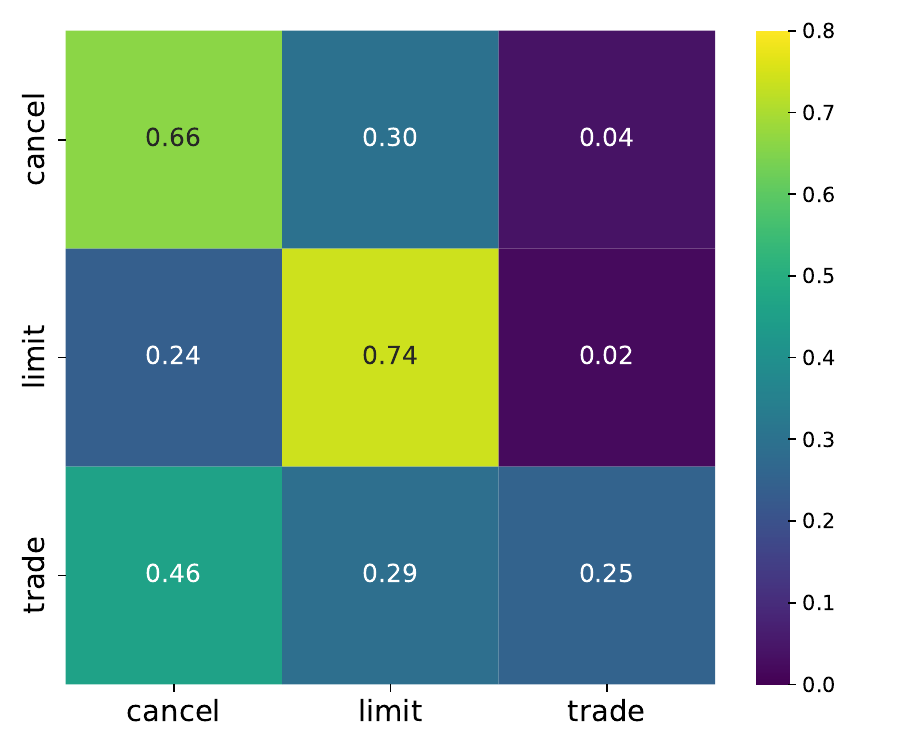}
        \caption{DQR}
        \label{fig:sub3}
    \end{subfigure}
    \caption{Transition matrix of events on simulated markets vs historical data. The rows of the matrix represent the conditional probabilities of observing a specific event type given the type of the previous event (intra-sides).}
    \label{fig:three_side_by_side}
\end{figure}

Figure~\ref{fig:three_side_by_side} compares the transition matrices of events:
\begin{itemize}
    \item \textbf{Historical data:} The ground truth, showing distinct excitation patterns.
    \item \textbf{SAQR model:} Rows are uniform, indicating the absence of excitation.
    \item \textbf{DQR model:} Captures realistic excitation patterns, closely matching historical data.
\end{itemize}

As seen in Figure~\ref{fig:three_side_by_side}, the DQR model reproduces excitation patterns observed in historical data, validating the importance of including $\eta_{k-1}$ in the state space. In contrast, the SAQR and QR models fail to capture this key dynamic, resulting in uniform transition matrices, meaning absence of any event type interdependencies. Previous work by~\cite{wu2019queue} successfully reproduced these excitation patterns by developing a hybrid framework combining Hawkes processes with the QR model. The DQR framework achieves comparable results capturing these dynamics by incorporating historical event information in the state space.

\subsubsection{Intraday seasonality}
Another limitation of the QR model is its uniform treatment of market behavior across different times of the day. In practice, markets exhibit distinct activity regimes throughout the trading session, with heightened activity during opening and closing periods and calmer phases around lunch break, as documented in~\cite{bodor2024stylized}. To capture these intraday patterns, we incorporate temporal information into the model's state space.

To capture intra-day seasonality, we included the \textit{hour of the last event} $h_k$ as an additional feature in the state space $x_k$. This transforms $x_k$ for each event $e_k$ into:
\[
x_k = [q_k, h_k],
\]
where $q_k$ is the size of the queue immediately before the event, and $h_k$ is the current hour of the day (as a categorical feature).

\begin{figure}[H]
\center
\includegraphics[width=0.6\textwidth, trim=0 0 0 23, clip]{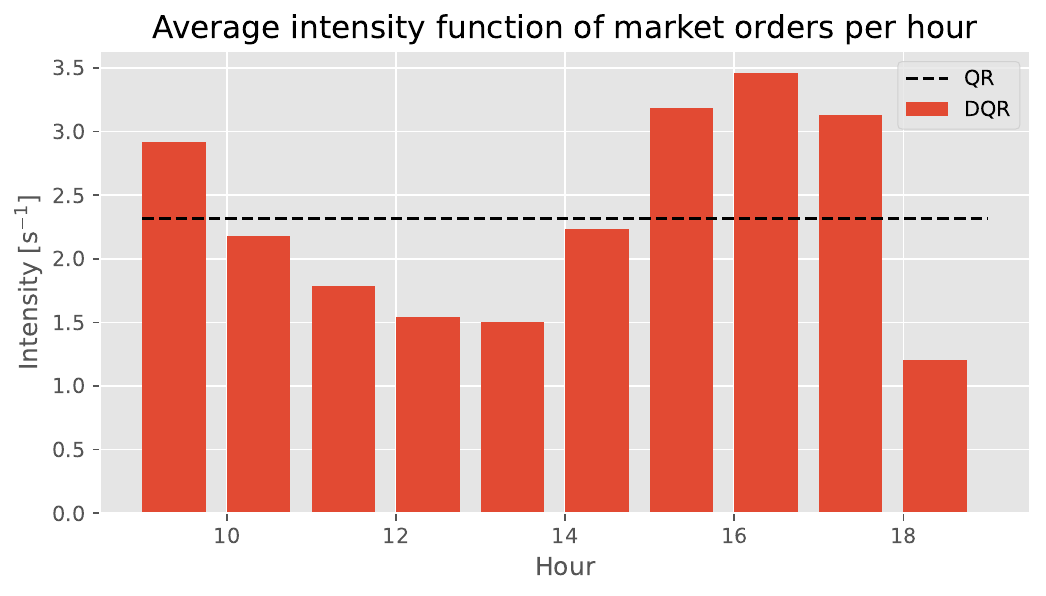}

\caption{Comparison of DQR (with $x_k = [q_k, h_k]$) and QR trade order arrival intensities across trading hours, averaged over queue sizes.}

\label{fig:intraday_seasonality}
\end{figure}

Including \( h_k \) enables the model to learn intra-day patterns, such as variations in event intensities across trading hours (e.g., heightened activity at european markets open and close). Figure~\ref{fig:intraday_seasonality} illustrates how incorporating \( h_k \) improves the ability of the DQR model to reproduce observed intra-day seasonality in market activity.

\subsubsection{Impact of Feature Enrichment on Model Performance}
To further validate the impact of enriching the state space \( x_k \), we evaluated the performance of the Deep Queue-Reactive (DQR) model across three key metrics:
\begin{itemize}
    \item \textbf{Log-likelihood of the model:} A higher log-likelihood indicates better alignment of the model's predictions with observed market events.
    \item \textbf{Balanced accuracy of next-event prediction:} The ability of the model to correctly classify the type of the next event (\( \eta_k \)).
    \item \textbf{Relative difference in time to the next event:} Measures how well the model predicts the timing of the next event compared to the actual observed values.
\end{itemize}

\begin{figure}[H]
    \centering
    \includegraphics[width=0.6\textwidth]{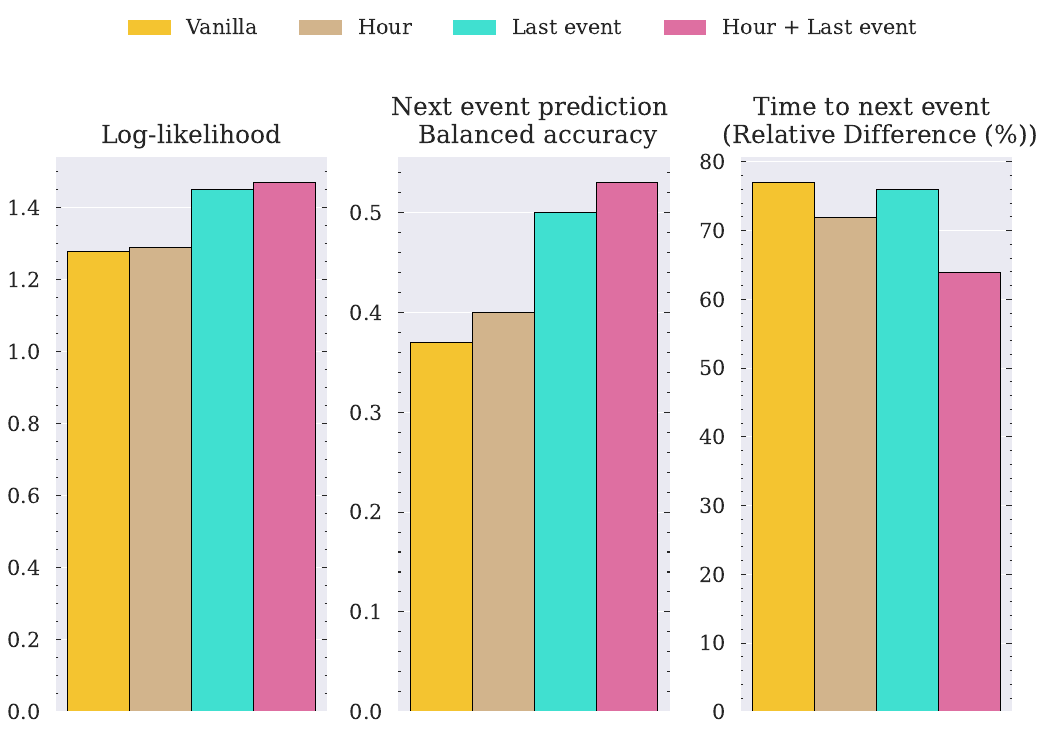}
    \caption{Model performance comparison across different feature sets: (1) Vanilla QR model (\( x_k = q_k \)), (2) \( x_k = [q_k, h_k] \), (3) \( x_k = [q_k, \eta_{k-1}] \), (4) \( x_k = [q_k, h_k, \eta_{k-1}] \). Metrics: log-likelihood (higher is better), balanced accuracy of next-event prediction (higher is better), and relative difference in time to next event (lower is better).}
    \label{fig:excitation_performance}
\end{figure}

The results, shown in Figure~\ref{fig:excitation_performance}, demonstrate the performance improvements achieved by incorporating additional features into \( x_k \), such as the hour of the last event (\( h_k \)) and the type of the last event (\( \eta_{k-1} \)).

\subsubsection*{Analysis of Results}

\textit{Model Likelihood (Left Panel):}
Using queue size $q_k$ alone as the baseline feature, we observe how additional features enhance the model's performance. Adding the hour of the last event ($h_k$) improves the log-likelihood by capturing intra-day seasonality patterns. Incorporating the type of the last event ($\eta_{k-1}$) yields further improvement by accounting for event excitation effects. The combination of both $h_k$ and $\eta_{k-1}$ leads to the highest log-likelihood, suggesting these features provide complementary information about order flow dynamics.

\textit{Balanced Accuracy (Middle Panel):}
Starting from the baseline model with queue size alone, we see systematic improvements in next-event type prediction with each feature addition. While $h_k$ captures time-of-day effects on order flow patterns, $\eta_{k-1}$ shows particularly strong impact by introducing event sequencing information. Combining both features yields the highest balanced accuracy, leveraging both temporal patterns and event dependencies.

\textit{Relative Difference in Time to the Next Event (Right Panel):}
The baseline model provides a reference point for inter-event time prediction. Adding $h_k$ significantly reduces the prediction error by incorporating intra-day seasonality, showing particular effectiveness in timing prediction. The addition of $\eta_{k-1}$ maintains this improvement in timing accuracy, while the combined model achieves the lowest relative difference by leveraging both the temporal structure from $h_k$ and the event sequencing from $\eta_{k-1}$.\\

The empirical results demonstrate the DQR framework's capacity to capture complex market dynamics through contextual features. The enhanced performance achieved by incorporating both $h_k$ and $\eta_{k-1}$ enables the model to reflect key market properties such as intra-day seasonality and event-type dependencies. Not only does this enriched state space improve the simulation quality, but it also enhances the model's likelihood, validating the incorporation of additional market features. This extension of the original framework maintains a balance between model sophistication and interpretability, preserving the underlying point-process structure while accommodating richer market dynamics.

These results suggest that the DQR model's ability to reproduce market properties depends on the features included in the state space $x$, with each addition potentially capturing new aspects of market behavior. This observation motivates further exploration of richer state spaces. In the next section, we present a generalization of this framework that incorporates a broader set of market features.

\section{Multidimensional Deep Queue-Reactive Model}

We have introduced the Queue-Reactive (QR) and Deep Queue-Reactive (DQR) models as frameworks for describing event arrivals in a single queue. QR models capture how different event types (limit, cancel, and market) depend on the queue size, while DQR generalizes this dependency to more complex state representations (e.g., additional market features, historical dependencies) through parametric (neural network-based) intensities. Although these models offer valuable insights, they share two critical limitations:

\begin{enumerate}
    \item \textbf{Independence Across Queues:} Both QR and DQR consider each queue or price level in isolation. In practice, however, different price levels within a limit order book interact with one another. For example, activity at the best bid can influence behavior at the best ask, and conditions at deeper levels can propagate to the top of the book. Ignoring these cross-level dependencies limits the realism and explanatory power of the model.

    \item \textbf{Absence of Order Size Modeling:} Neither QR nor DQR explicitly models the distribution of order sizes. Empirical findings \cite{bodor2024novel} indicate that order sizes play a pivotal role in shaping limit order book dynamics. By omitting order sizes, we lose an essential dimension of market complexity, potentially reducing the model’s ability to accurately reflect observed phenomena.
\end{enumerate}

To address these issues, we propose the \emph{Multidimensional Deep Queue-Reactive (MDQR)} model. The MDQR model extends the QR framework in three principal ways: (i) it generalizes the state space from only queue size $q_k$ to a general state space $x_k$ and models intensity of arrival of events using deep learning (as introduced in the DQR model), (ii) it treats the entire limit order book as a single multidimensional entity, thereby capturing cross-level interactions, and (iii) it incorporates order sizes into the probabilistic formulation, allowing a more comprehensive representation of market events.

\begin{figure}[h!]
\centering
\includegraphics[width=\textwidth]{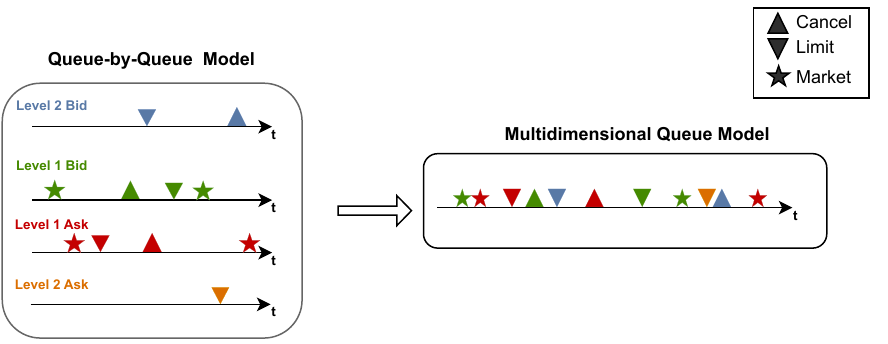}
\caption{A conceptual illustration of the transition from a single-queue model (left) to the multidimensional MDQR framework (right). In the single-queue scenario, events occur at one level and are of three types (L, C, M). In the multidimensional scenario, multiple levels on both sides of the order book are considered together, resulting in a larger event space and enabling the model to capture cross-level interactions and dependencies.}
\label{fig:multi_DQR_Queue_representation}
\end{figure}

Figure~\ref{fig:multi_DQR_Queue_representation} illustrates the conceptual shift from a single-queue setting to a multidimensional one. On the left-hand side, the traditional view considers a single queue receiving three types of events (limit, cancel, market) for each queue. On the right-hand side, the MDQR model expands this perspective to multiple price levels, each capable of receiving any of the three event types. This results in a richer event space with \(3 \times 2n\) distinct category-level combinations if \(n\) price levels per side are considered (where positive indices represent ask-side levels and negative indices represent bid-side levels). By unifying these levels within a single modeling framework, the MDQR model can directly capture how conditions and events at one level influence those at another.

To formalize this approach, consider a sequence of events:
\[
\mathcal{E} = \{e_k\}_{k=1}^N,
\]
where each event
\[
e_k = (\eta_k, \ell_k, \Delta t_k, s_k, x_k)
\]
is characterized by:
\begin{itemize}
    \item \(\eta_k \in \{\mathrm{L}, \mathrm{C}, \mathrm{M}\}\): The event category (limit, cancel, market).
    \item \(\ell_k \in \mathcal{P} = \{-K, \ldots, -1, 1, \ldots, K\}\): The price level at which the event occurs. Positive integers correspond to ask-side levels and negative integers to bid-side levels.
    \item \(\Delta t_k = t_k - t_{k-1}\): The inter-arrival time since the previous event.
    \item \(s_k\): The size of the order associated with the event.
    \item \(x_k\): A state vector that may include information from multiple levels and sides, as well as other relevant features such as historical dependencies or market indicators.
\end{itemize}

The joint likelihood of the observed event sequence factorizes into two components:

\begin{equation}
\mathcal{L}(\theta \mid \mathcal{E}) \;=\; \left(\prod_{k=1}^{N} p(\eta_k, \ell_k, t_k \mid x_k;\theta)\right) \times \left(\prod_{k=1}^{N} p(s_k \mid \eta_k, \ell_k, t_k, x_k;\theta)\right)
\end{equation}

The first part of this factorization governs the timing and category-level characteristics of events. As in the QR and DQR models, we assume a conditional Poisson structure for event arrivals. Define an intensity function \(\lambda^{(\eta,\ell)}(x_k;\theta)\) for each event category \(\eta\) and level \(\ell\). The total intensity of observing any event at any level is:
\[
\Lambda(x_k;\theta) \;=\; \sum_{\eta \in \{\mathrm{L},\mathrm{C},\mathrm{M}\}} \sum_{\ell \in \mathcal{P}} \lambda^{(\eta,\ell)}(x_k;\theta)
\]

The probability of an event of type $\eta_k$, at level $\ell_k$ occurring at time \(t_k\) given the state \(x_k\) is then:

$$
p(\eta_k, \ell_k, t_k \mid x_k;\theta) \;\propto\; e^{-\Lambda(x_k;\theta)\,\Delta t_k} \, \lambda^{(\eta_k,\ell_k)}(x_k;\theta)
$$

and the joint probability over all events for this part of the model is:
\[
\prod_{k=1}^{N} p(\eta_k, \ell_k, t_k \mid x_k;\theta) \;\propto\; \prod_{k=1}^{N} e^{-\Lambda(x_k;\theta)\,\Delta t_k} \, \lambda^{(\eta_k,\ell_k)}(x_k;\theta).
\]

This expression generalizes the DQR framework to a multidimensional setting, where the state \(x_k\) can integrate information from multiple price levels, capturing cross-level dependencies and a richer set of market signals. 

For model calibration, we minimize the negative log-likelihood, which yields a more tractable optimization problem:
\begin{equation}
\label{eq:log_lik_MDQR}
\ell_{\lambda}(\theta) = \sum_{k=1}^N \Lambda(x_k;\theta)\Delta t_k - \log \lambda^{(\eta_k,\ell_k)}(x_k;\theta)
\end{equation}
where $\ell_{\lambda}(\theta)$ represents the objective function to be minimized to calibrate the arrival intensities model.

The second component of the likelihood,
\[
\prod_{k=1}^{N} p(s_k \mid \eta_k, \ell_k, t_k, x_k;\theta),
\]
models the distribution of order sizes given the event type, level, and state. This factor is independent of the timing and category-level intensities, allowing it to be modeled separately. In a subsequent section, we will focus on how to specify and estimate this size distribution.

In summary, the MDQR model combines a multidimensional representation of the order book with an integrated modeling of order sizes. By capturing cross-level interactions and explicitly incorporating size distributions, MDQR offers a more comprehensive and realistic framework for analyzing and simulating limit order book dynamics than previous single-queue models.

\subsection{Modeling the Order Size Distribution}

Recall from the MDQR model formulation that the joint likelihood of the observed event sequence factors into two components:
\[
\mathcal{L}(\theta \mid \mathcal{E}) \;=\; \left(\prod_{k=1}^{N} p(\eta_k, \ell_k, t_k \mid x_k;\theta)\right) \times \left(\prod_{k=1}^{N} p(s_k \mid \eta_k, \ell_k, t_k, x_k;\theta)\right)
\]
The first term, as discussed, governs the timing, level, and category of events and has been handled in the previous section. The second term models the distribution of order sizes conditioned on the event characteristics and state.

To specify \( p(s_k \mid \eta_k, \ell_k, t_k, x_k;\theta) \), we require a flexible approach that can capture the diversity of size distributions observed in practice. One might consider parametric distributions such as Lognormal, Gamma, or Weibull, with parameters conditioned on \((\eta_k, \ell_k, x_k)\). However, order size distributions often display substantial skewness and can vary dramatically depending on the event type and available liquidity. For example, sizes associated with certain event types may be effectively bounded by the remaining liquidity at that level, making it difficult for a single parametric family to fit all scenarios satisfactorily.

An alternative is to employ mixture models, such as mixtures of Gaussians. While these can approximate a wide range of distributions, they introduce additional complexity in model selection (e.g., choosing the number of components) and parameter estimation. Computational overhead and convergence stability can also become challenging, especially in settings with many conditioning variables.

Instead, a more straightforward yet effective approach is to discretize the order size space into a set of predefined classes and model the distribution as a categorical probability over these classes. This approach transforms the continuous modeling problem into a multi-class classification task, allowing us to leverage standard techniques such as cross-entropy loss minimization. The discretization can be achieved by splitting the range of order sizes into bins or quantiles, ensuring that each class corresponds to a distinct range of sizes.

In principle, if order sizes vary widely, one could define a suitable binning strategy, for example by grouping them into quantiles to maintain balanced class frequencies. This would yield a flexible, data-driven set of classes that adapt to the empirical distribution of observed order sizes.

In our specific dataset, however, over 99.9\% of observed order sizes are less than 200. This natural concentration near smaller values simplifies the approach significantly. We can assign each integer size from 1 up to 200 to its own class, eliminating the need for binning or quantile-based grouping. As a result, the order size modeling reduces to predicting a categorical distribution over 200 classes (200 possible order size). A neural network can then be trained, taking as input \((\eta_k, \ell_k, x_k)\), to produce the probability of each discrete order size class. This neural network’s parameters are estimated by minimizing the cross-entropy loss across all observed events.

Should a different dataset exhibit a larger or more dispersed range of order sizes, binning into quantiles or equal-width intervals could be employed to keep the number of classes manageable and to maintain balanced representation across them. Thus, the approach is flexible and can be adapted to various data characteristics.

This classification-based strategy offers a balance between flexibility and practicality. It does not rely on restrictive parametric forms, avoids the complexity of mixture models, and naturally accommodates state-dependent and event-dependent patterns through the neural network's capacity to learn non-linear relationships. This approach provides a suitable framework for modeling order sizes in the MDQR model. The likelihood maximization for this component translates to minimizing the cross-entropy loss:

\begin{equation}
\label{eq:MDQR_sizes_loss}
\ell_s(\theta) = -\sum_{k=1}^N \sum_{c=1}^C y_{k,c} \log \hat{p}_c(s_k \mid \eta_k, \ell_k, x_k;\theta)
\end{equation}

where $\hat{p}_c$ denotes the model's predicted probability for class $c$, $y_{k,c}$ is the one-hot encoded ground truth (1 if $s_k$ belongs to class $c$, 0 otherwise), and $C$ is the total number of size classes ($C = 200$ in our case). The probabilities are conditioned on the event characteristics ($\eta_k, \ell_k$) and market state $x_k$.

\subsection{Data}
The data used to calibrate the MDQR model follows the framework established in Section~\ref{sec:Data1}, with one key modification in the temporal sampling approach. While the previous analysis computed $\Delta t_k$ between events within individual queues, the present study considers time intervals between all order book events, providing a more comprehensive view of market dynamics. The bucketing methodology and simulation paradigm remain consistent with the earlier framework, ensuring continuity in our analytical approach.

\subsection{Features}
The model incorporates a diverse set of features capturing both instantaneous market state and historical market dynamics. Let us denote the limit order book state at time $t_k$ as follows: for any level $i$, let $q_i(t_k)$ represent the queue size, where $i \in \{-M,\ldots,-1,1,\ldots,M\}$, with negative indices corresponding to bid levels and positive indices to ask levels. Thus, $q_{-1}(t_k)$ and $q_1(t_k)$ represent the best bid and best ask queue sizes respectively.

For each price level $i$, we maintain a categorical variable $e_i(t_k)$ representing the type of the last event (limit order, cancellation, or market order) that occurred at that level. The spread $s(t_k)$ is computed as the difference between the best ask and best bid prices:

\begin{equation*}
    s(t_k) = p_i(t_k) - p_j(t_k), \quad \text{where} \quad i = \min\{l > 0 : q_l(t_k) > 0\}, \quad j = \max\{l < 0 : q_l(t_k) > 0\}
\end{equation*}

To capture order flow imbalances across different time horizons, we define the trade imbalance feature $TI_\tau(t_k)$ over a window $\tau$ as:

\begin{equation*}
    TI_\tau(t_k) = \frac{\sum_{t \in [t_k-\tau, t_k]} V^b(t) - \sum_{t \in [t_k-\tau, t_k]} V^a(t)}{\sum_{t \in [t_k-\tau, t_k]} V^b(t) + \sum_{t \in [t_k-\tau, t_k]} V^a(t)}
\end{equation*}

where $V^b(t)$ and $V^a(t)$ represent the traded volumes at the bid and ask sides respectively at time $t$. We consider multiple time horizons $\tau \in \mathcal{H} = \{20\textrm{s}, 1\textrm{min}, 5\textrm{min}, 15\textrm{min}\}$
to capture both short-term and long-term market tendencies.

To enhance the model's ability to learn from categorical features, each categorical variable (event types $e_i(t_k)$) is processed through an embedding layer of dimension 2, leveraging the relatively low cardinality of these features while allowing the model to learn meaningful representations of event sequences.

\begin{table}[h]
\centering
\caption{Summary of Model Features and Their Characteristics}
\label{tab:features}
\begin{tabular}{llll}
\hline
\textbf{Feature Type} & \textbf{Description} & \textbf{Preprocessing} & \textbf{Properties} \\
\hline
\multicolumn{4}{l}{\textit{Numerical Features}} \\
Queue Sizes ($q_i(t_k)$) & Level-wise volumes & Log transformation & $i \in \mathcal{P}$ \\
Spread ($s(t_k)$) & Price differential & None & Measured in ticks \\
Trade Imbalance ($TI_\tau(t_k)$) & Trades Imbalance & None & $\tau \in \mathcal{H}$ \\
\hline
\multicolumn{4}{l}{\textit{Categorical Features}} \\
Last Event Type ($e_i(t_k)$) & Previous events & Embedding (dim=2) & Cardinality: 3 \\
Hour ($h(t_k)$) & Current hour & Embedding (dim=2) & Cardinality: 9 \\
\hline
\end{tabular}
\end{table}

\subsection{MDQR model Architecture}

The MDQR framework extends the architectural principles established for the DQR model (Section~\ref{sec:DQR_architecture}), adapting them to accommodate multiple price levels. The intensity prediction network employs a Multi-Layer Perceptron with two hidden layers of dimensions 256 and 64 neurons respectively. The output layer, now expanded to dimension 30, encompasses the complete event space across all monitored price levels—three possible event types (limit orders, cancellations, and market orders) for each of the ten price levels (five bid and five ask levels). The network maintains hyperbolic tangent activation functions throughout its hidden layers.

For the order size model, we preserve the core architectural elements while adjusting the network dimensions to accommodate the broader scope of market dynamics. The model retains Tanh activations in its hidden layers, with a Softmax activation in the output layer to generate appropriate probability distributions.

Table~\ref{tab:architecture_comparison} presents the architectural specifications for both the intensity and order size models.

\begin{figure}[H]
    \centering
    \includegraphics[width=0.7\textwidth]{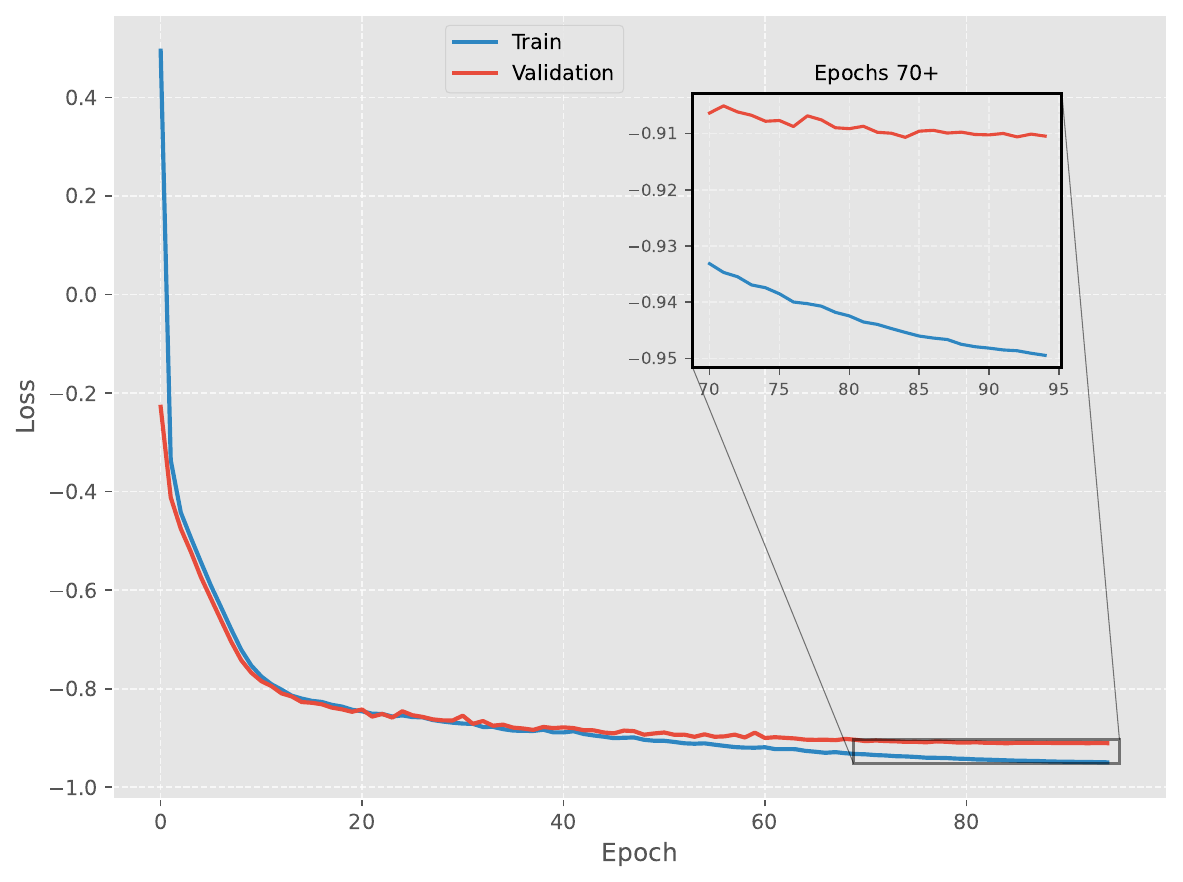}
\caption{Training and validation negative log-likelihood loss evolution for the order intensity model. The main plot shows the full training trajectory, while the inset focuses on epochs 70-94 to highlight convergence behavior.}
    \label{fig:learning_curve}
\end{figure}

Figure~\ref{fig:learning_curve} shows the learning curves of the neural network trained to predict order intensities, displaying the negative log-likelihood loss for both training and validation sets. The model exhibits stable convergence, and training was stopped after the validation loss showed no improvement for approximately 10 epochs around epoch 85.

\begin{table}[H]
\centering
\caption{Architectural Specifications of MDQR Components}
\label{tab:architecture_comparison}
\begin{tabular}{lll}
\hline
\textbf{Specification} & \textbf{Intensity Model} & \textbf{Order Size Model} \\
\hline
Input Dimension & 25 & 27 \\
Hidden Layers & [256, 64] & [256, 64] \\
Output Dimension & 30 & 200 \\
Hidden Activations & Tanh & Tanh \\
Output Activation & ReLU & Softmax \\
Loss Function & Eq.~(\ref{eq:log_lik_MDQR}) & Eq.~(\ref{eq:MDQR_sizes_loss}) \\
Optimizer & Adam & Adam \\
Learning Rate Range & $10^{-5}$ to $10^{-3}$ & $10^{-5}$ to $10^{-3}$ \\
\hline
\end{tabular}
\end{table}

%
%
%
%
%
%
%
%
%
%
%

\subsection{Results}
To evaluate the performance of the proposed simulator, we adopt a comprehensive validation approach that examines multiple aspects of market behavior. Our analysis focuses on three key areas: the model's ability to reproduce established market stylized facts, its capacity to capture complex patterns in order book dynamics, and its predictive power for market movements.

\subsubsection{Market Impact}

One of the most important use cases of limit order book (LOB) simulators is leveraging them as environments to train reinforcement learning (RL) agents or backtest execution strategies. A critical aspect of these environments is their ability to accurately replicate the market's response to an agent's actions, particularly concerning market impact. Market impact refers to the effect that executing a large order has on price evolution, whether the order is executed instantaneously or through a specific splitting strategy \citep{gatheral2010no, bouchaud2018trades}.

\begin{figure}[H]
    \centering
    \includegraphics[width=0.7\textwidth]{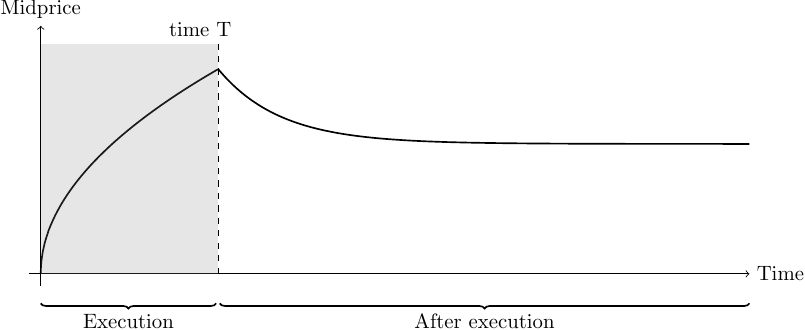}
\caption{Average shape of price impact.
During its execution, a buy metaorder drives the price upward, reaching a peak impact upon completion ($t = T$). Following the cessation of buying pressure, the price reverts sharply. However, a residual impact often remains observable well beyond the execution period and may, in some cases, persist permanently, often called permanant merket impact. Figure from \cite{coletta2022conditional}.}
    \label{fig:MI_theoretical}
\end{figure}

Empirical studies have consistently observed that during a time-weighted average price (TWAP) execution of a large buy order, the price typically increases in a concave manner until reaching a peak—known as the temporary market impact—and then decreases in a convex fashion, eventually settling at a permanent market impact level. This characteristic price trajectory is depicted in Figure~\ref{fig:MI_theoretical}. Notable studies supporting these findings include \citep{zarinelli2015beyond}, \citep{bacry2015market} and \citep{bouchaud2018trades}.

Another extensively studied phenomenon is the relationship between the magnitude of market impact and the executed quantity. The prevailing consensus is that the maximum impact does not increase linearly with the size of the executed quantity but follows a square-root law. This implies that if $I(Q)$ represents the market impact for a quantity $Q$, then $I(Q) \propto \sqrt{Q}$. This relationship has been documented in various studies, including \citep{bucci2019crossover}, \citep{sato2024does}, and \citep{gabaix2003theory}. \\


\begin{figure}[h]
    \centering
    \includegraphics[width=0.7\textwidth]{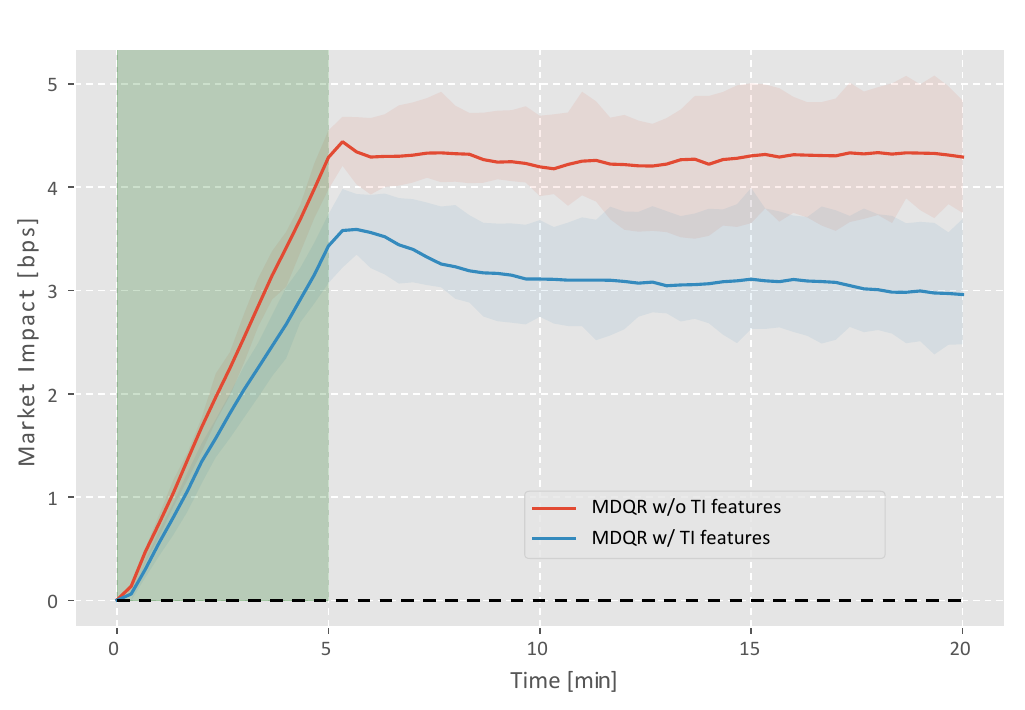}
    \caption{Market impact profiles for two models: one using only queue sizes and the other incorporating trade imbalance features. Results are based on a TWAP agent executing 100\% of the average traded volume within a 5-minute window. The order is split into 10 child orders executed at a rate of 30 seconds.}
    \label{fig:MI_with_and_without_TI}
\end{figure}

The market impact profiles displayed in Figure~\ref{fig:MI_with_and_without_TI} reveal distinct behavioral patterns between models with and without Trade Imbalance (TI) features. The model without TI features exhibits a linear increase in price impact during the execution phase, followed by a plateau phase where the impact stabilizes. In contrast, the model incorporating TI features demonstrates a more nuanced and theoretically grounded behavior: it shows a concave-like progression during the execution phase, followed by a gradual relaxation period that stabilizes approximately 10 minutes after execution completion.

This characteristic shape aligns with the empirical findings of~\cite{coletta2022conditional}, who documented similar market impact dynamics. Furthermore, the observed pattern closely matches theoretical predictions~\citep{zarinelli2015beyond, bouchaud2018trades} and corresponds to the theoretical market impact profile presented in Figure~\ref{fig:MI_theoretical}, though with a subtle concavity in the execution phase and a relaxation stabilizing at a modest 20\% reduction---a magnitude whose market-specificity cannot be verified without meta-trade data. The inclusion of trade imbalance features thus enhances the MDQR model's ability to capture realistic market dynamics, particularly in reproducing the well-documented non-linear nature of price impact and its subsequent relaxation phase.

The improved realism in market impact modeling demonstrates that incorporating trade imbalance features allows the model to better capture the complex interplay between order flow and price formation processes in financial markets. 

\begin{figure}[H]
    \centering
\includegraphics[width=0.47\textwidth, trim=0cm 0cm 0cm 1cm, clip]{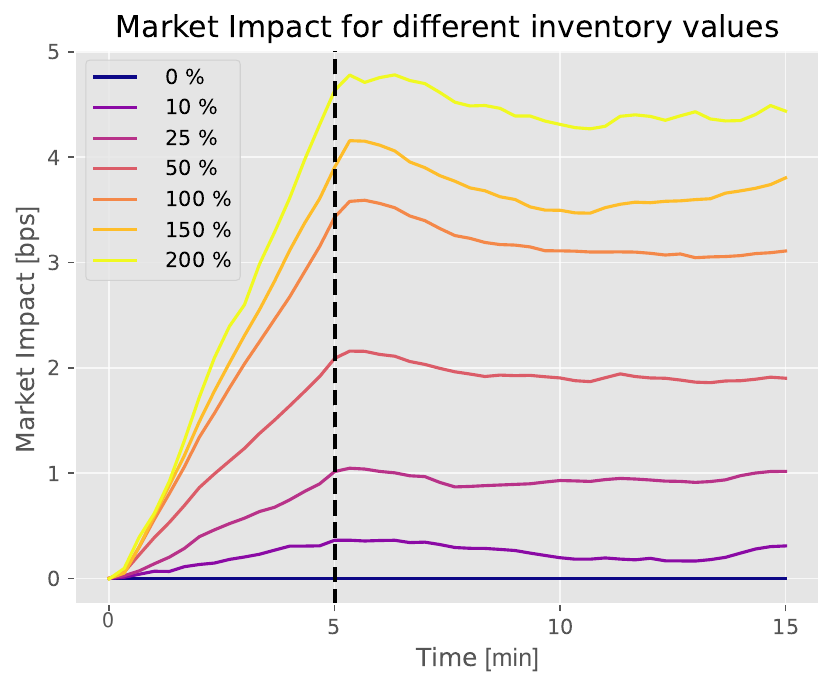}
    \hfill
    \includegraphics[width=0.51\textwidth]{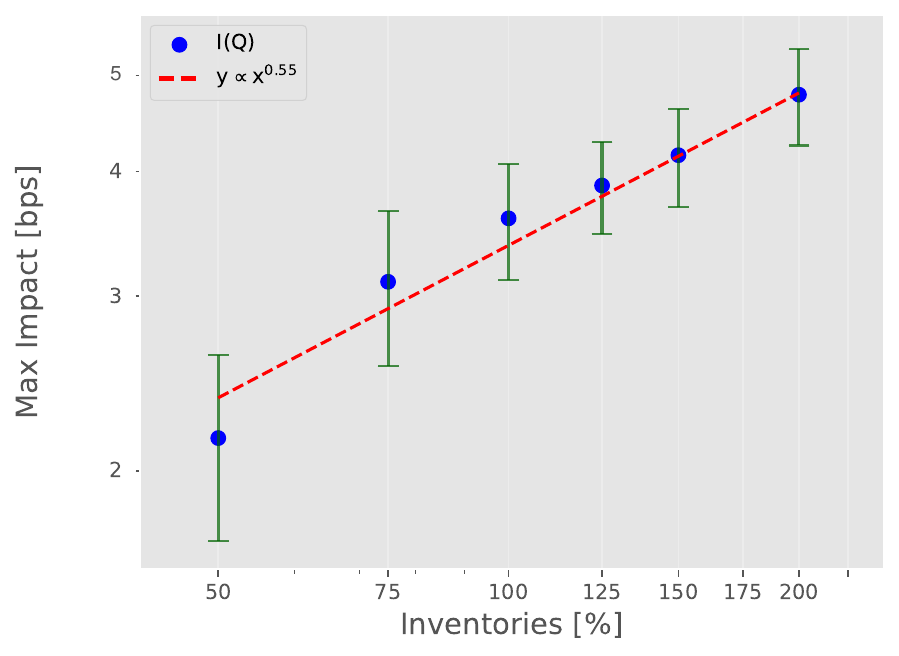}
    \caption{Average price change for different inventories (left) and maximum market impact ($I(Q)$) as a function of initial inventory size  (in log-log scale) with the exponent best  fit (right). Vertical lines indicate the 95\% confidence itervals.}
    \label{fig:square_root_law}
\end{figure}

Furthermore, Figure~\ref{fig:square_root_law} provides additional insights into the model's behavior under varying trading volumes. The left panel illustrates the evolution of market impact profiles for different initial quantities $q_0$ (quantity to buy at $T=5$ minutes), expressed as percentages of the average 5-minute trading volume in the Bund futures market (approximately 4,000 lots, which is notably consistent with typical liquidity levels in this major fixed-income futures market). As the inventory size increases from 0\% to 200\% of the average traded volume in 5 minutes in the Bund market, we observe a systematic increase in market impact magnitude while maintaining similar temporal dynamics. The right panel provides empirical validation of the square-root law of market impact: plotting the maximum impact against inventory size on logarithmic scales shows a fit to a $x^{0.55}$ power law with $R^2 = 0.89$, indicating that the model reproduces this market characteristic.

The improved realism in market impact modeling demonstrates that incorporating trade imbalance features allows the model to better capture the complex interplay between order flow and price formation processes in financial markets. This advancement represents a significant step toward more accurate representation of market microstructure dynamics in quantitative trading models.

\subsubsection{Excitation between different sides}

\begin{figure}[H]
    \centering
    \begin{subfigure}[b]{0.32\textwidth}
        \includegraphics[width=\textwidth,clip]{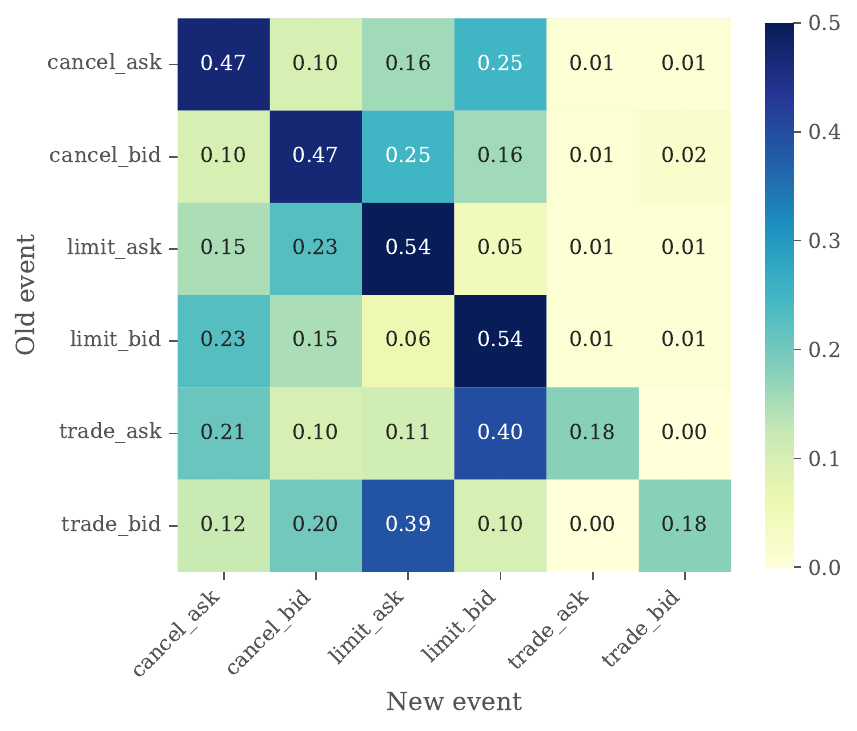}
        \caption{Real}
        \label{fig:sub1}
    \end{subfigure}
    \hfill 
    \begin{subfigure}[b]{0.32\textwidth}
        \includegraphics[width=\textwidth,clip]{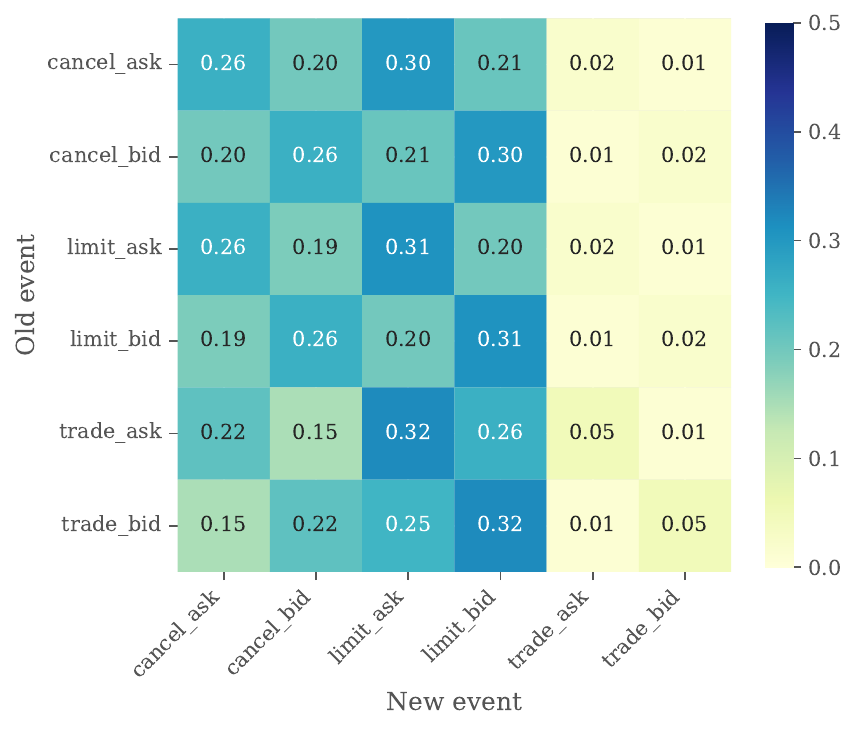}
        \caption{QR \& SAQR}
        \label{fig:sub2}
    \end{subfigure}
    \hfill 
    \begin{subfigure}[b]{0.32\textwidth}
        \includegraphics[width=\textwidth,clip]{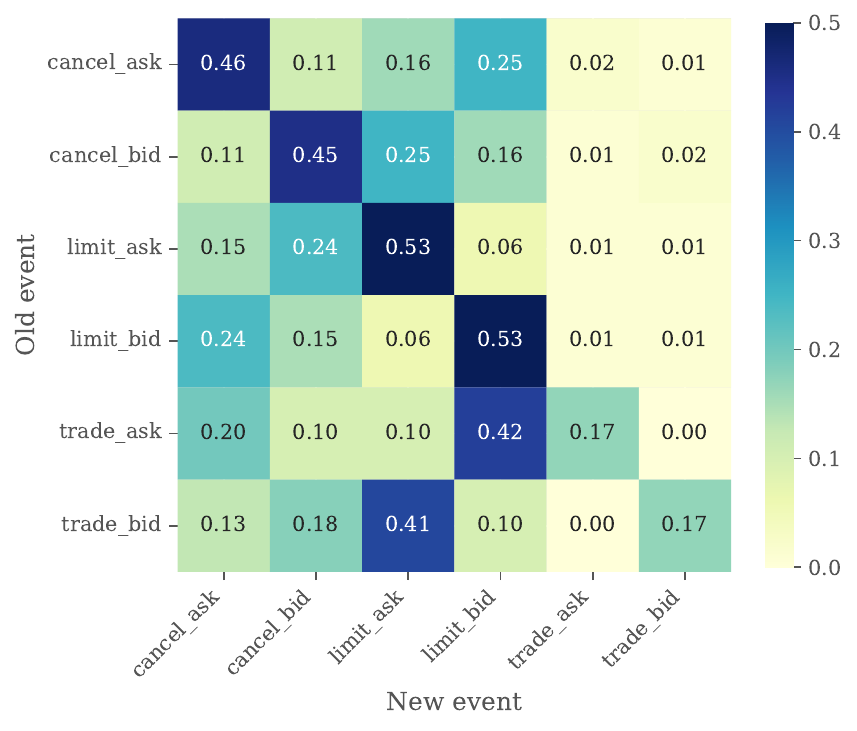}
        \caption{MDQR}
        \label{fig:sub3}
    \end{subfigure}
    \caption{Transition matrix of events on simulated markets vs historical for events on the best prices.}
    \label{fig:exitation_sides}
 \end{figure}

A key aspect of market microstructure modeling is understanding the temporal dependencies between different order book events. The transition matrix $\mathbb{P}(E_{k+1}|E_k)$, where $E_k$ represents the event type at step $k$, captures these dependencies by measuring the probability of transitioning from one event type to another. One of the primary motivations for relaxing the independence assumption between queues is to better model the interactions between different price levels, particularly between the best ask and best bid prices.

Figure~\ref{fig:exitation_sides} presents a comparative analysis of these transition matrices across three scenarios: historical data, SAQR model, and MDQR model. In the SAQR model, we observe that the rows of the transition matrix exhibit similar patterns, indicating minimal event-specific excitation effects. This similarity suggests that the SAQR model, with its independent queue assumption, fails to capture the complex interactions present in the order book.
In contrast, the MDQR model, leveraging its ability to model different price levels simultaneously, successfully reproduces transition probabilities that closely match the historical patterns. This improved accuracy is particularly evident in the cross-excitation effects between ask and bid events, demonstrating the model's capability to capture the subtle interdependencies that characterize real market dynamics.

\subsection{Mid-Price Prediction}

While the primary objective of our framework is to generate realistic order book dynamics, evaluating its predictive capabilities provides additional validation of its ability to capture price formation mechanisms. We adopt the methodology of~\cite{zhang2019deeplob} to assess short-term price prediction performance, with modifications to account for the specific characteristics of the highly liquid Bund futures market.

The prediction task involves classifying directional price movements over a horizon of $k=500$ events, where this longer horizon reflects the higher liquidity of the Bund market compared to equity markets. For each time point $t_0$, we compute:

\begin{equation*}
    m_{-} = \frac{1}{k}\sum_{j=0}^{k-1} x_j^{\text{mid}}, \quad m_{+} = \frac{1}{k}\sum_{j=1}^{k} x_j^{\text{mid}}, \quad r = m_{+} - m_{-}
\end{equation*}

where $x_j^{\text{mid}}$ represents the mid-price $j$ events before $t_0$ for $j<0$, at $t_0$ for $j=0$, or $j$ events after $t_0$ for $j>0$. Price movements are classified as upward when $r > \Delta r$, downward when $r < -\Delta r$, and stationary otherwise. The threshold $\Delta r$ is set to 20 basis points to achieve balanced class distributions in the Bund market context.

For simulator-based predictions, we generate 500 forward events starting from each observed order book state. The predicted movement is determined by computing $\hat{m}_+$ from these simulated trajectories and comparing $\hat{m}_+ - m_-$ against the threshold $\Delta r$.

The DeepLOB benchmark model is calibrated using one month of limit order book data, with a two-day validation period and three-day test period. The remaining data is used for training, with early stopping implemented when the validation loss shows no improvement for 10 consecutive epochs. The model is trained on the last $T=500$ order book states, which consist of five levels of the order book on each side. A learning rate of $1 \times 10^{-5}$ and a batch size of 1024 are used during training. Table~\ref{tab:price_movement} summarizes the proportion of price movements across training, validation, and test sets, showing reasonable balance between the three classes across all datasets.

\begin{table}[h]
\centering
\caption{Proportion of Price Movements in Datasets}
\label{tab:price_movement}
\begin{tabular}{lccc}
\toprule
\textbf{Price Move} & \textbf{Train} & \textbf{Validation} & \textbf{Test} \\
\midrule
Stationary & 0.38 & 0.31 & 0.37 \\
Up        & 0.31 & 0.36 & 0.32 \\
Down      & 0.31 & 0.33 & 0.31 \\
\bottomrule
\end{tabular}
\end{table}

\begin{table}[H]
\centering
\caption{Comparison of Models: Balanced Accuracy and F1 Score.}
\label{tab:model_comparison}
\begin{tabular}{lcc}
\toprule
\textbf{Model} & \textbf{Balanced Accuracy} & \textbf{F1 Score} \\
\midrule
DeepLOB & 0.54 $\pm$ 0.01 & 0.56 $\pm$ 0.01 \\
QR      & 0.56 $\pm$ 0.01 & 0.55 $\pm$ 0.02 \\
SAQR    & 0.58 $\pm$ 0.03 & 0.58 $\pm$ 0.03 \\
MDQR    & 0.63 $\pm$ 0.02 & 0.62 $\pm$ 0.02 \\
\bottomrule
\end{tabular}
\end{table}

Table~\ref{tab:model_comparison} presents the predictive performance of various models in forecasting mid-price movements. While the DeepLOB architecture, specifically designed for price prediction, achieves a baseline performance with balanced accuracy of 0.54, the generative models show competitive or superior results. The QR model matches this performance despite its simpler architecture, while the SAQR demonstrates modest improvements in both metrics. The MDQR framework achieves the highest performance across both metrics, with a balanced accuracy of 0.63 and F1 score of 0.62, suggesting that its enhanced ability to capture cross-price level dependencies contributes to improved predictive power. These results are particularly noteworthy given that price prediction is not the primary objective of these generative models.


\subsection{Other Stylized Facts}



\subsubsection{Queue sizes distribution}

\begin{figure}[H]
    \centering
    \includegraphics[width=0.54\textwidth]{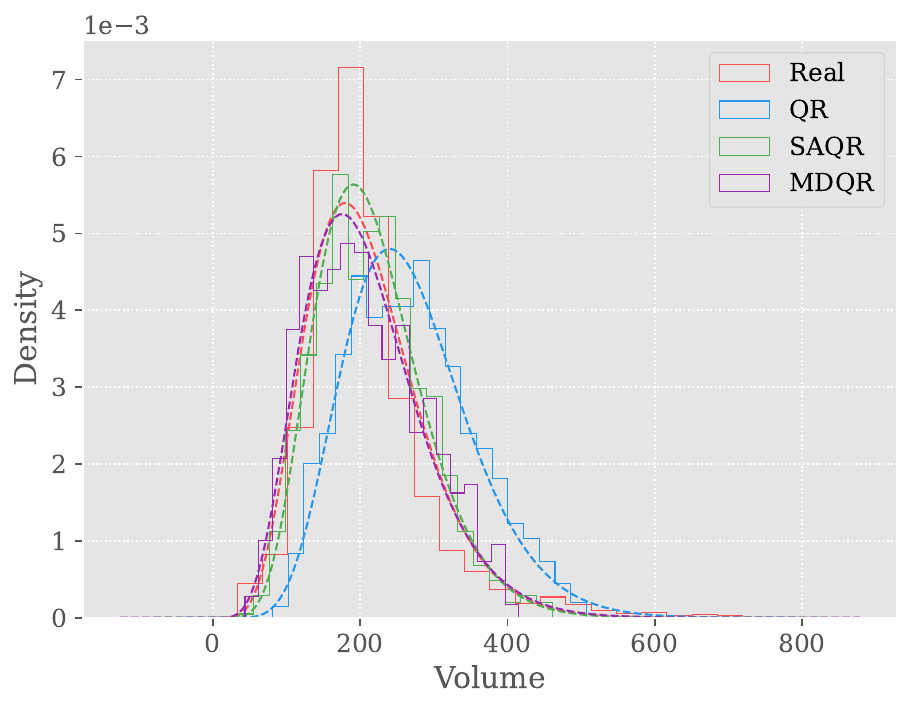}
    \hfill
    \includegraphics[width=0.42\textwidth]{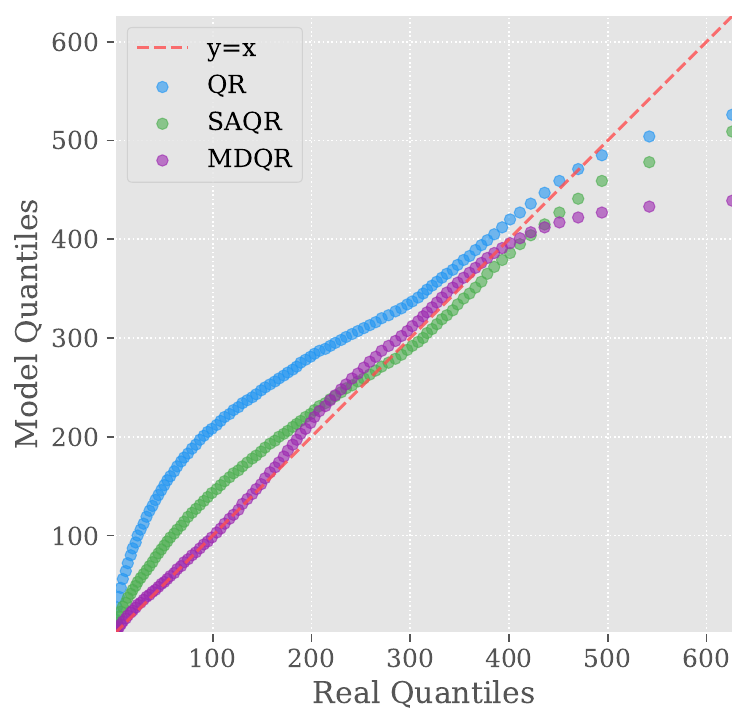}
  \caption{Comparison of real and simulated order book queue sizes. \textbf{Left:} Distribution of best ask volumes from real data versus different QR model variants, alongside a gamma distribution fit. \textbf{Right:} Q-Q plot comparing simulated market queue sizes against real queue sizes, illustrating the alignment between model predictions and actual data.}
    \label{fig:order_book_comparison}
\end{figure}

\begin{table}[h]
\centering
\begin{tabular}{lcc}
\hline
Model & $\alpha$ & $1/\beta$ \\
\hline
Real & $1.35 \pm 0.18$ & $183.44 \pm 31.79$ \\
QR & $3.08 \pm 0.13$ & $83.43 \pm 3.06$ \\
SAQR & $1.91 \pm 0.06$ & $110.49 \pm 3.14$ \\
MDQR & $1.30 \pm 0.13$ & $172.00 \pm 12.26$ \\
\hline
\end{tabular}
\caption{Characteristics of gamma fit for shape and scale parameters (mean $\pm$ std)}
\label{tab:gamma_fit}
\end{table}

The distributional properties of queue sizes provide crucial insights into the models' ability to capture fundamental market dynamics. Figure~\ref{fig:order_book_comparison} presents a comparative analysis of best ask volumes across different model specifications. The MDQR model demonstrates notable accuracy in reproducing the empirical distribution, particularly evident in the left panel where the simulated density closely tracks the historical pattern. This alignment is further validated through the Q-Q plot analysis, showing remarkable consistency between model-generated and empirical quantiles.


A particularly noteworthy aspect is the close correspondence between the empirical distribution and a gamma distribution fit, aligning with documented stylized facts in fixed income markets~\citep{bodor2024stylized}. The gamma distribution, characterized by its probability density function:

\[
f(x;\alpha,\beta) = \frac{\beta^\alpha x^{\alpha-1}e^{-\beta x}}{\Gamma(\alpha)}
\]

\noindent where $\alpha$ represents the shape parameter and $\beta$ the rate parameter (with scale = $1/\beta$), provides a natural framework for modeling queue sizes. Table~\ref{tab:gamma_fit} shows that the MDQR model closely reproduces the empirical distribution's parameters, with shape ($\alpha = 1.30 \pm 0.13$) and scale ($1/\beta = 172.00 \pm 12.26$) values closely matching those of historical data ($\alpha = 1.35 \pm 0.18$, $1/\beta = 183.44 \pm 31.79$). This alignment suggests that our model effectively captures the underlying properties of queue size formation, while simpler models like QR and SAQR show notable deviations in their distributional parameters.\\

\begin{figure}[H]
    \centering
    \includegraphics[width=0.85\textwidth]{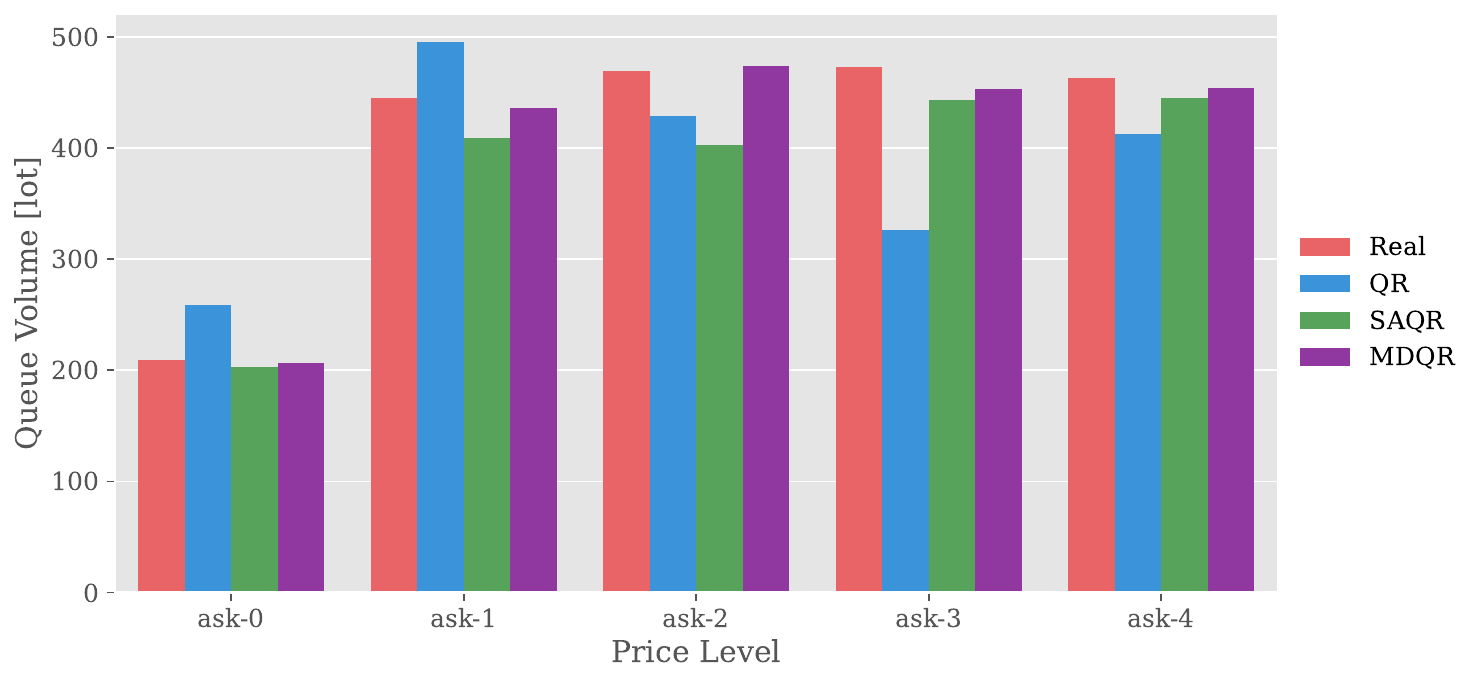}
    \caption{Average LOB volumes for the first five levels at the ask side.}
    \label{fig:all_levels_queue_sizes}

\end{figure}

Figure~\ref{fig:all_levels_queue_sizes} extends this analysis to deeper price levels, comparing average posted volumes between models and historical data. The MDQR framework maintains its superior performance across different price levels, consistently producing volume patterns that closely match empirical observations. This broader success in reproducing volume characteristics suggests that the model effectively captures not just the dynamics at the best quotes, but also the complex interactions that shape liquidity provision across the order book's depth.

\subsubsection{Correlation between Order Book Queues}

\begin{figure}[h]
    \centering
    \begin{subfigure}[b]{0.45\textwidth}
        \centering
        \includegraphics[width=\textwidth]{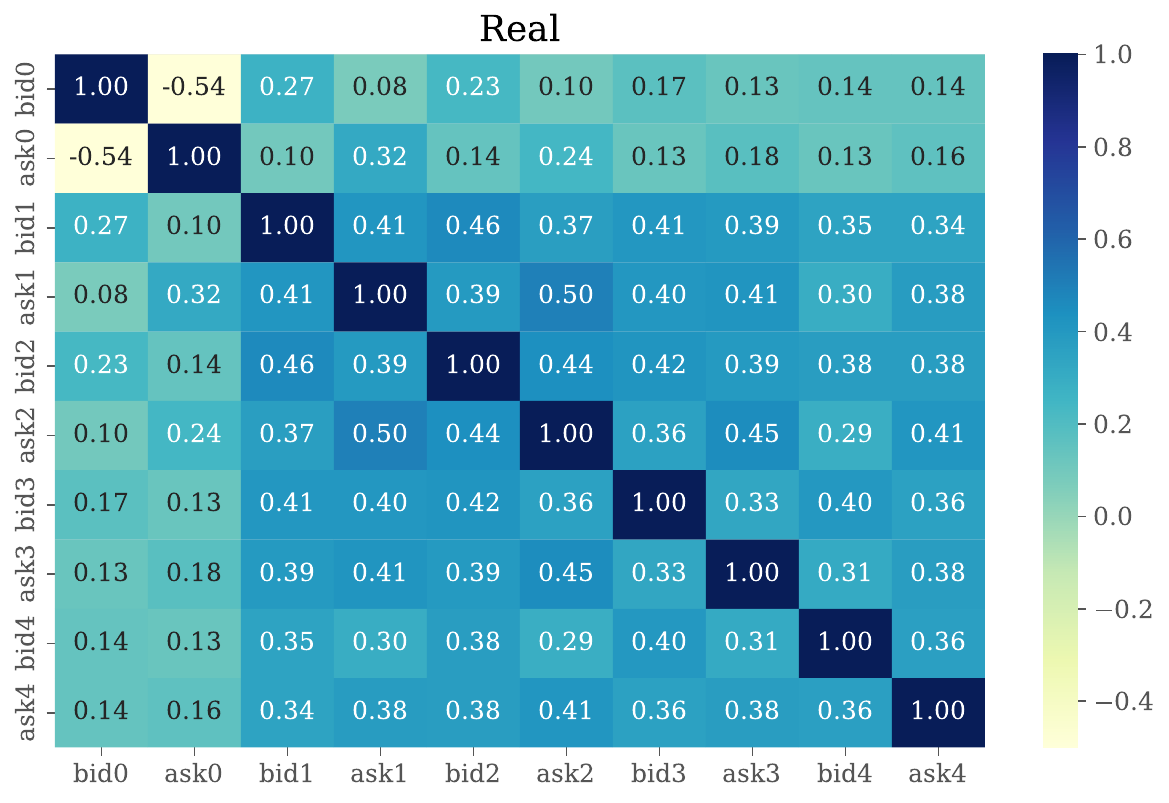}
        \caption{Real}
        \label{fig:subfig1}
    \end{subfigure}
    \hfill
    \begin{subfigure}[b]{0.45\textwidth}
        \centering
        \includegraphics[width=\textwidth]{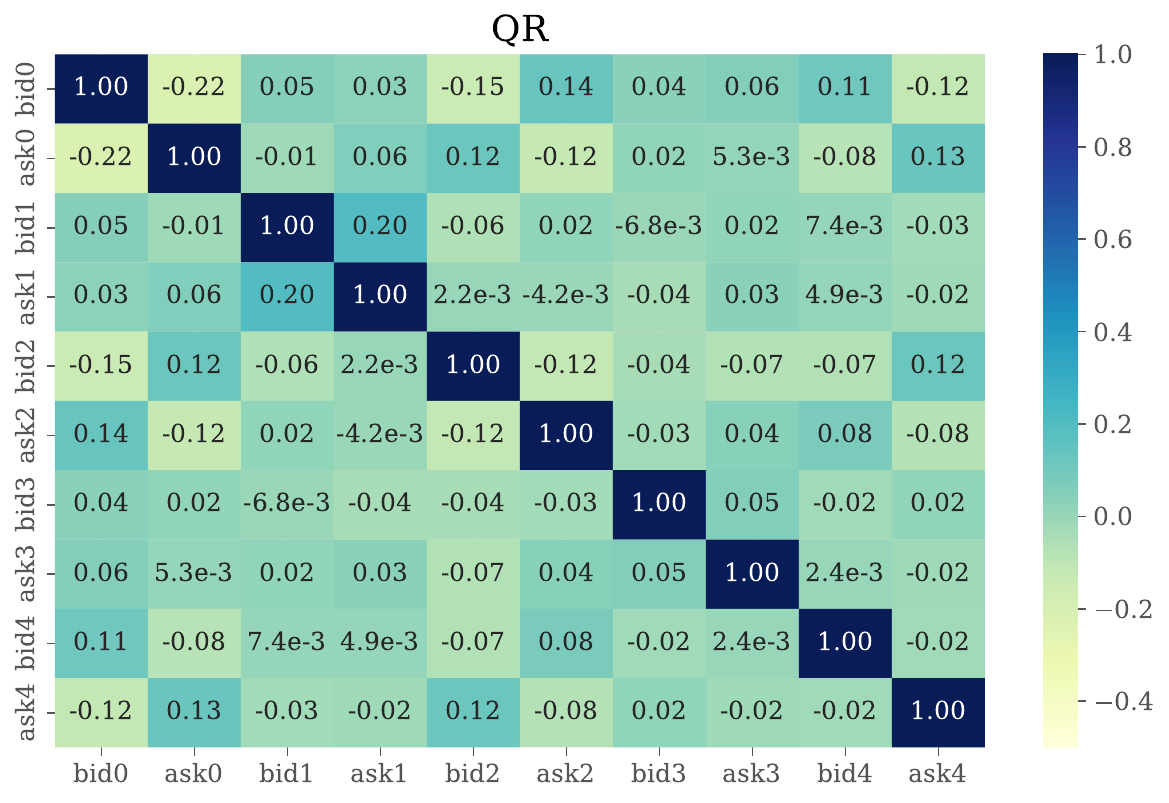}
        \caption{QR}
        \label{fig:subfig2}
    \end{subfigure}

    \vskip\baselineskip

    \begin{subfigure}[b]{0.45\textwidth}
        \centering
        \includegraphics[width=\textwidth]{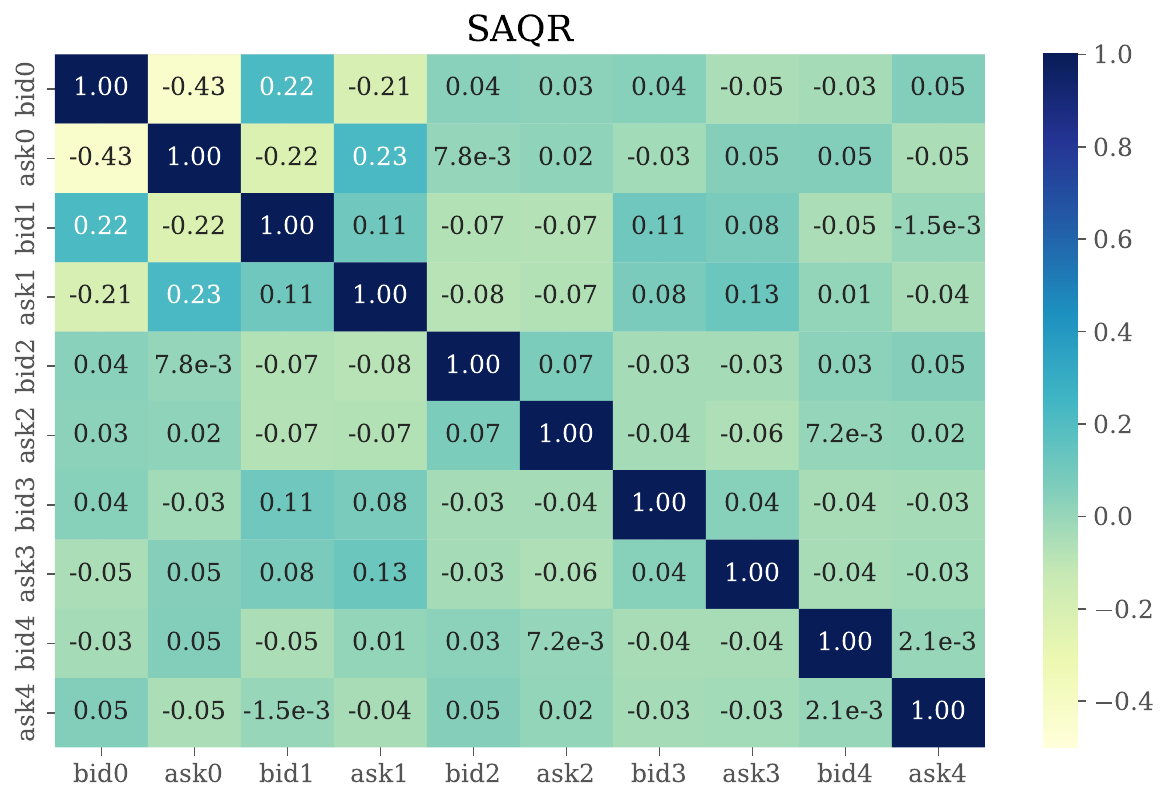}
        \caption{SAQR}
        \label{fig:subfig3}
    \end{subfigure}
    \hfill
    \begin{subfigure}[b]{0.45\textwidth}
        \centering
        \includegraphics[width=\textwidth]{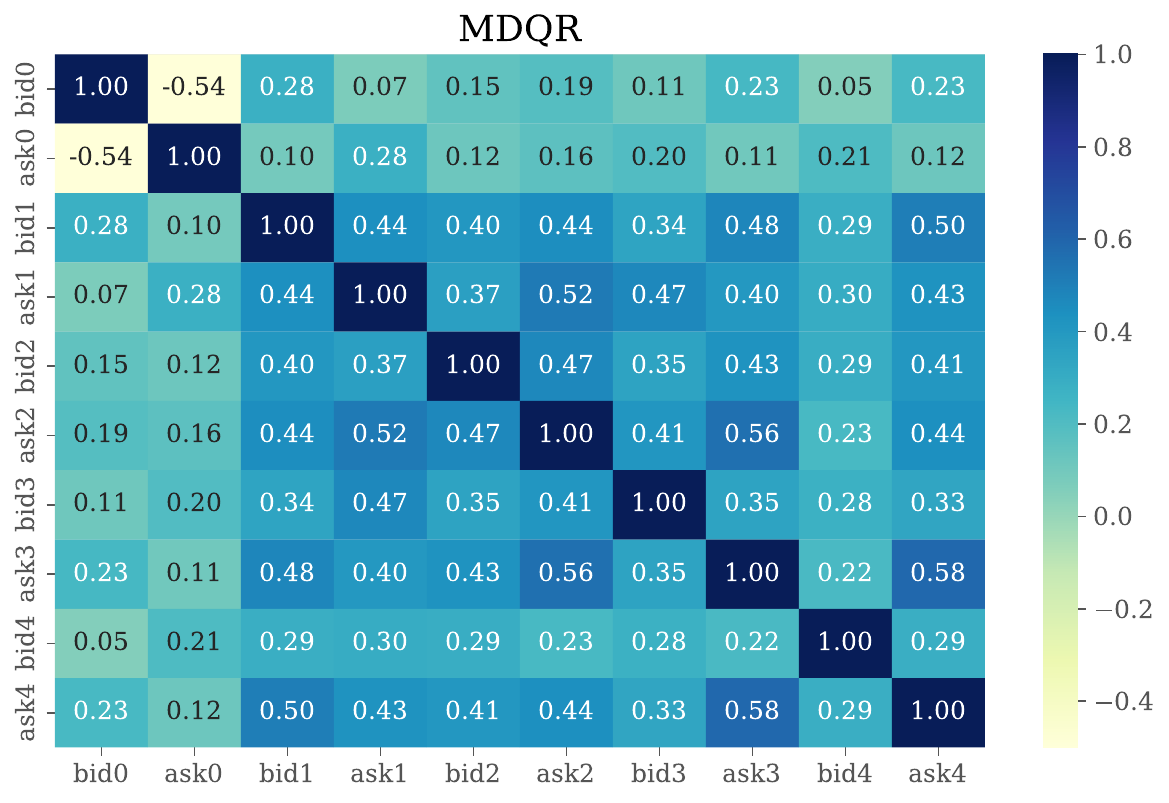}
        \caption{MDQR}
        \label{fig:subfig4}
    \end{subfigure}

  \caption{Correlation matrices of queue volumes across different price levels (5 bid and 5 ask levels) for historical data (Real) and three model variants (QR, SAQR, and MDQR). Each entry $(i,j)$ represents the Pearson correlation coefficient between volumes at price levels $i$ and $j$, where negative indices denote bid levels and positive indices denote ask levels.}

    \label{fig:correlation_volumes}
\end{figure}

The correlation structure between volumes at different price levels provides key insights into the models' ability to capture market microstructure dynamics. Figure~\ref{fig:correlation_volumes} presents correlation matrices between queue volumes across bid and ask levels, comparing empirical patterns with those generated by the QR, SAQR, and MDQR models.

The empirical correlation matrix (Figure~\ref{fig:correlation_volumes}a) reveals several notable patterns. First, there exists a strong negative correlation (-0.54) between the best bid and ask queues, reflecting the natural asymmetry in order flow: periods of high selling pressure often coincide with depleted buying interest and vice versa. Second, we observe significant positive correlations between queues on the same side of the book, with values typically ranging from 0.3 to 0.5, indicating coordinated liquidity provision behaviors.

The QR model (Figure~\ref{fig:correlation_volumes}b) fails to capture these essential relationships. Most strikingly, it underestimates the negative correlation between best bid and ask (-0.22 versus -0.54 in empirical data) and shows negligible correlations between queues at different price levels, with most values close to zero. This limitation stems from the model's independent treatment of individual queues.

While the SAQR model (Figure~\ref{fig:correlation_volumes}c) shows modest improvement in capturing the best bid-ask correlation (-0.43), it still struggles to reproduce the broader correlation structure, particularly the positive correlations between queues on the same side of the book. Most off-diagonal elements remain close to zero, indicating that the model fails to capture the interconnected nature of liquidity dynamics.

The MDQR model (Figure~\ref{fig:correlation_volumes}d) demonstrates markedly superior performance in reproducing the empirical correlation structure. It accurately captures both the negative correlation between best bid and ask (-0.54) and the positive correlations between same-side queues. The model generates realistic correlation patterns that extend beyond immediate price levels, suggesting its effectiveness in capturing the complex interplay of liquidity provision and consumption across the order book, all of this thanks to the MDQR ability to take into account states from multiples queues to simulate events.

\subsubsection{Distribution of Returns}

\begin{figure}[h]
    \centering
    \includegraphics[width=0.43\textwidth]{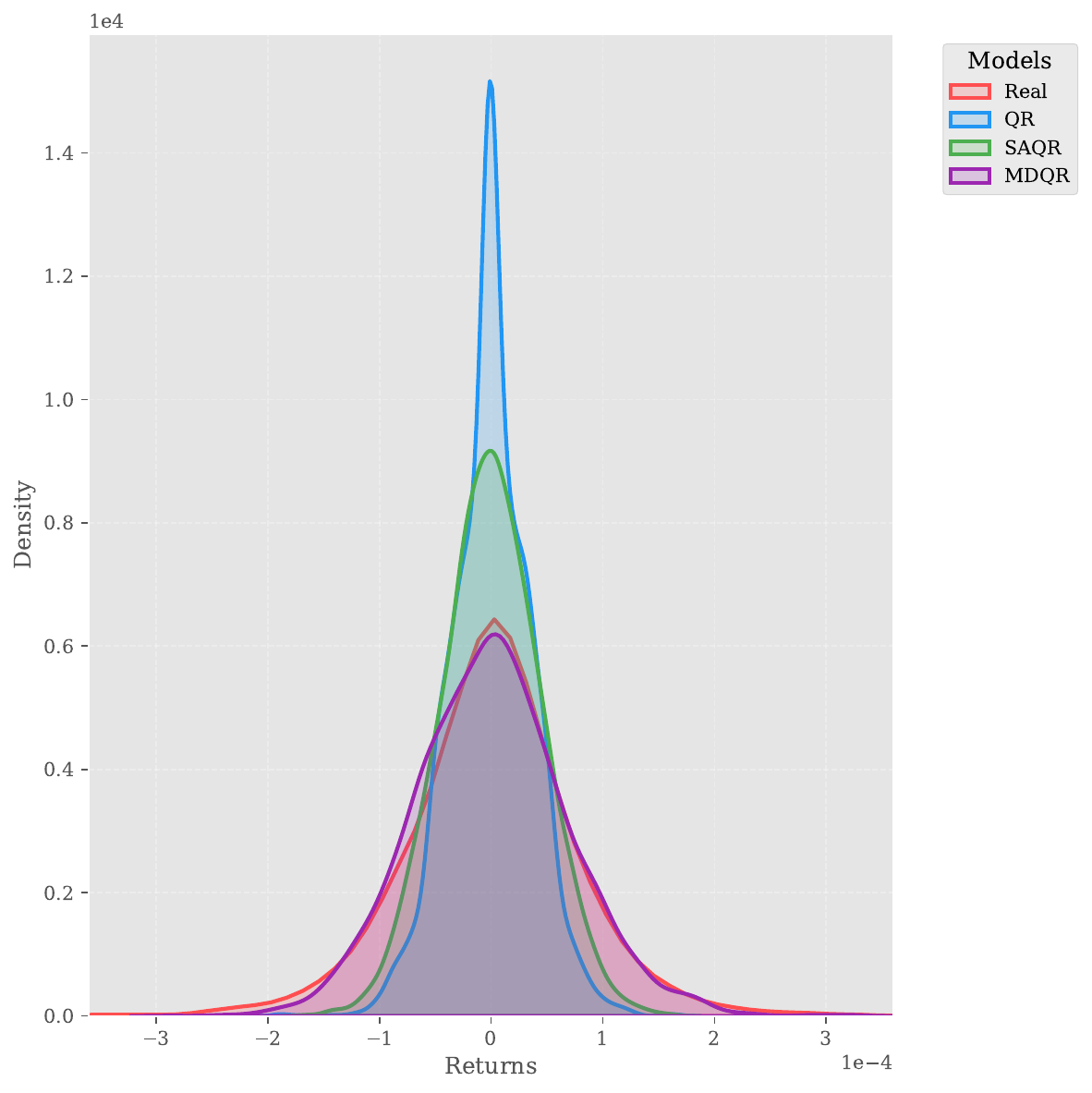}
    \hfill
    \includegraphics[width=0.43\textwidth]{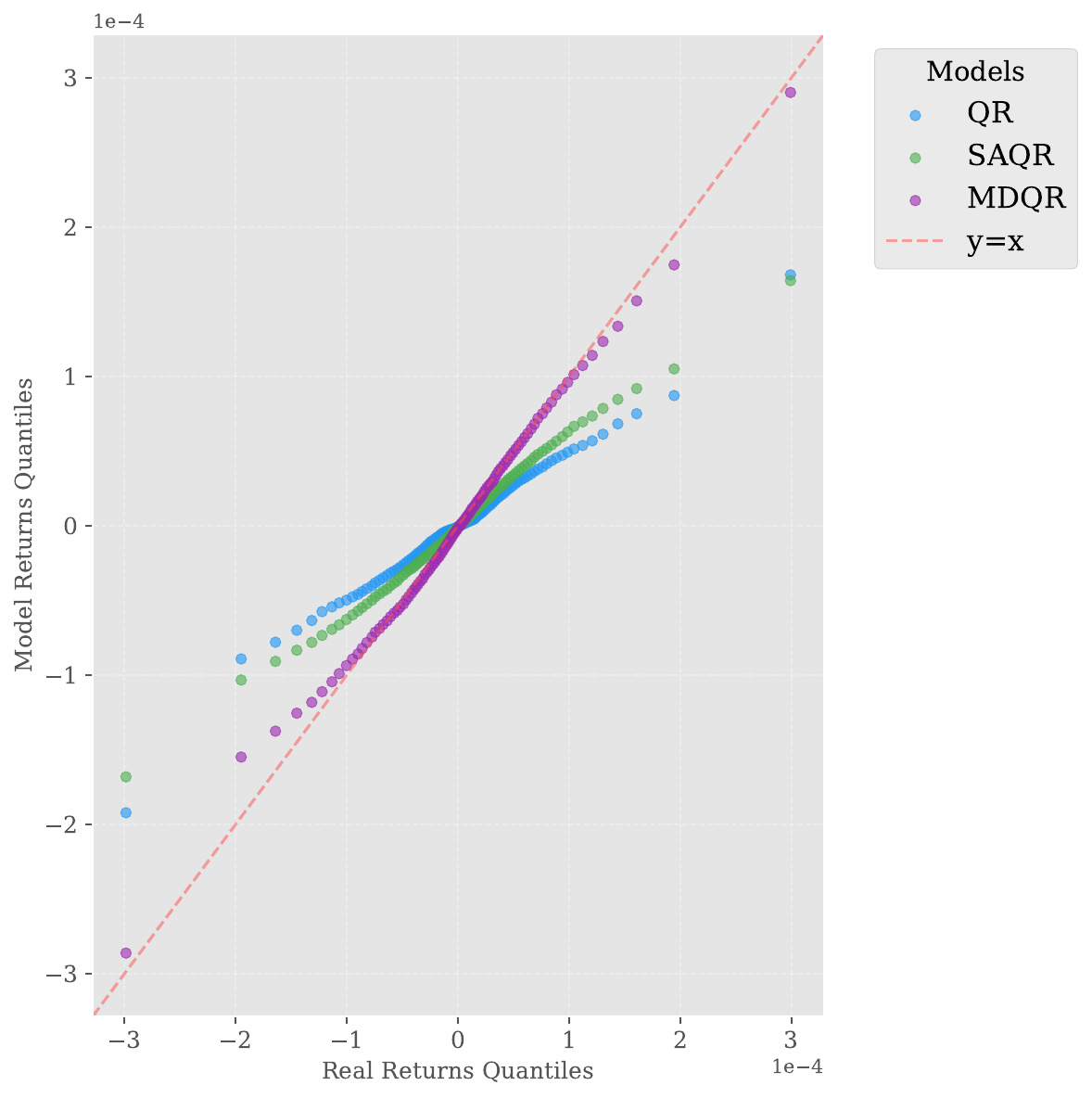}
    \caption{\textbf{Left:} Distribution of one-minute returns for real data and various model variants, highlighting differences in return profiles. \textbf{Right:} Q-Q plot comparing model-generated returns with real returns, showing the degree of alignment between simulated and actual data.}
    \label{fig:returns_comparison}

\end{figure}
  
The distributional properties of one-minute returns provide valuable insights into the models' capacity to capture market dynamics. Figure~\ref{fig:returns_comparison} presents both the probability density functions and quantile-quantile plots for the various model specifications. The MDQR model demonstrates superior performance in reproducing the empirical distribution of returns, particularly in capturing the characteristic heavy tails of the historical returns.

This improved distributional fit is further validated through the Q-Q plot analysis, where the MDQR quantiles closely align with the theoretical line $y=x$, indicating strong agreement between simulated and empirical distributions across the entire range of returns. In contrast, the QR model exhibits notable deviations, particularly in the tail regions. As discussed in~\cite{bodor2024novel}, these limitations stem from the QR model's inability to adequately capture rapid price movements due to its simplified queue dynamics.

The enhanced distributional accuracy of the MDQR model can be attributed to its more sophisticated treatment of cross-price level interactions and the incorporation of market-wide state variables, allowing it to better reflect the complex mechanisms driving price formation in real markets.

\subsubsection{Frequency of events}

\begin{figure}[H]
    \centering
    \includegraphics[width=0.9\textwidth]{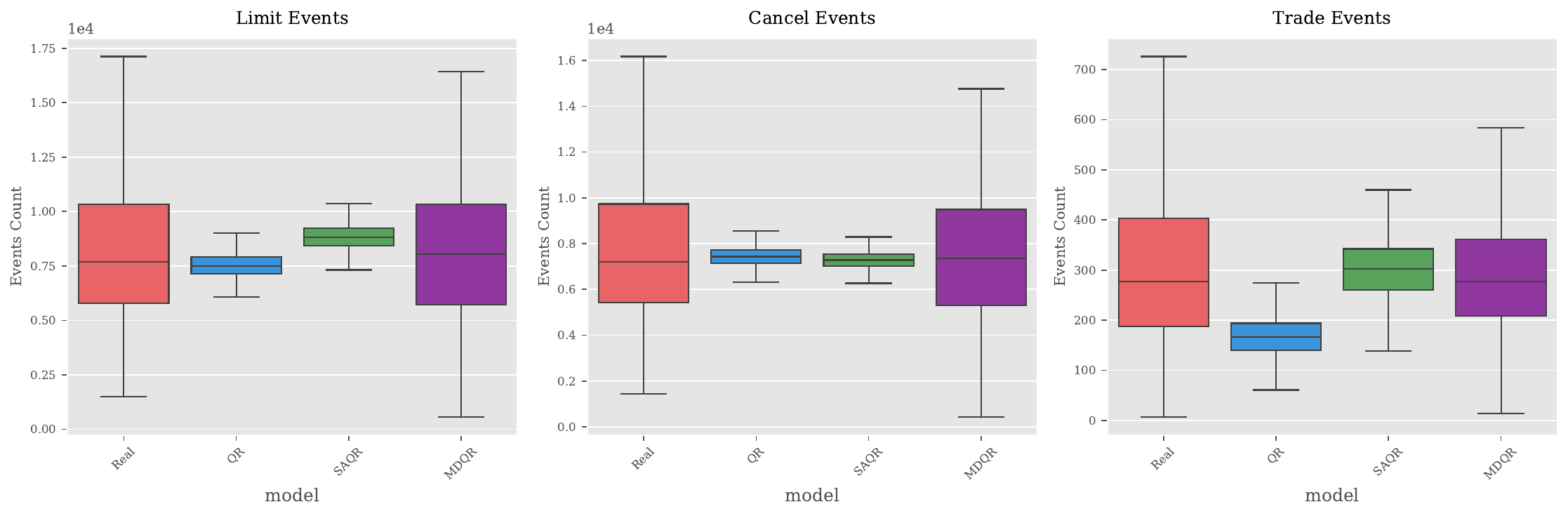}
    \caption{Distribution of event counts per 5-minute window across different model specifications. Box plots show the comparison between historical data (Real) and three model variants (QR, SAQR, and MDQR) for limit orders, cancellations, and trades.}
    \label{fig:frequ_events_count}
\end{figure}

\begin{figure}[H]
    \centering
    \includegraphics[width=0.9\textwidth]{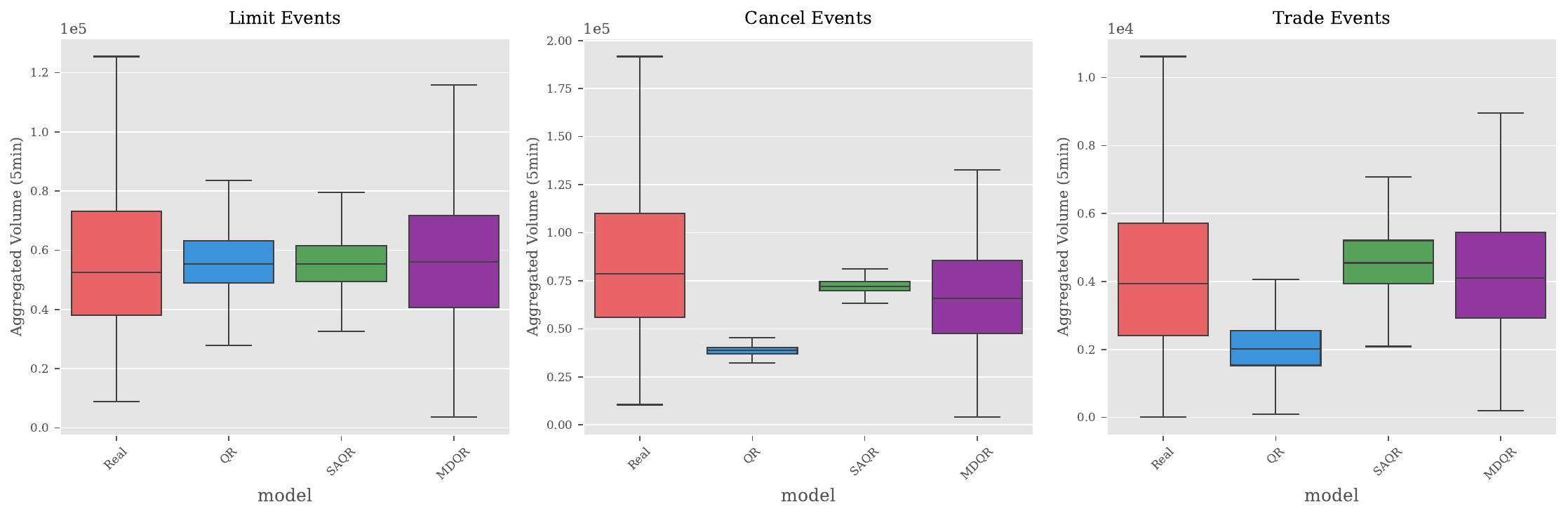}
    \caption{Distribution of aggregated volumes per 5-minute window across different model specifications. Box plots compare historical data (Real) with QR, SAQR, and MDQR models for limit orders, cancellations, and trades, showing both median values and dispersion patterns.}
    \label{fig:frequ_events_sum}
\end{figure}

Figures~\ref{fig:frequ_events_count} and~\ref{fig:frequ_events_sum} present a comprehensive comparison of event distributions across different model specifications through box plots, examining both event counts and aggregated volumes within 5-minute windows. The analysis reveals several key insights about the models' capabilities and limitations.

The QR model demonstrates accurate reproduction of average behavior for limit and cancel orders, both in terms of event counts and volumes. However, it shows significant discrepancy in modeling trade events. The SAQR extension successfully addresses this limitation, bringing the trade event distributions closer to historical patterns. This improvement highlights the importance of incorporating additional queue dependencies in the modeling framework.

Nevertheless, both QR and SAQR models, relying solely on queue sizes, exhibit a notable limitation: they fail to capture the true market variability. The observed standard deviations in these models are considerably smaller than those present in historical data, indicating an oversimplified representation of market dynamics.

The MDQR model addresses these shortcomings by not only reproducing average behaviors accurately but also capturing a wider range of market scenarios. The distributions more closely match the historical data across all event types, particularly in terms of variability. This enhanced capability stems from the model's ability to incorporate multiple price levels and their interactions, resulting in a more realistic representation of market microstructure dynamics.


\subsubsection{Simulation Time}
\begin{table}[H]
\centering
\begin{tabular}{lcc}
\hline
\textbf{Model} & \textbf{Inference} & \textbf{Daily} \\
& \textbf{Time (ms)} & \textbf{Generation (min)} \\
\hline
MDQR & $0.037 \pm 0.001$ & 0.92\\
LOBGAN \citep{sirignano2019universal} & $0.217 \pm 0.015$ & 5.43 \\
RNN \citep{hultin2023generative} & $1.152 \pm 0.061$ & 28.85 \\
WGAN \citep{cont2023simulation} & $0.144 \pm 0.001$ & 1.95 \\
\hline
\end{tabular}
\caption{Performance comparison of different architectures for limit order book modeling. Inference time is reported as mean $\pm$ standard deviation for generating a single event. Daily generation time represents the total computation time required to generate 1.5M events, equivalent, on average, to one trading day (9 AM to 6 PM) in the Bund Market.}
\label{tab:model_perf}
\end{table}

The computational efficiency of order book simulators plays a crucial role in various applications. While generating synthetic data might require a reasonable number of simulations, making computational speed less critical, reinforcement learning algorithms typically need billions of market events for each training run. Moreover, these simulations generally cannot be reused between different training iterations. This makes computational efficiency a crucial factor for practical implementation.


Table~\ref{tab:model_perf} presents a performance comparison across different limit order book modeling approaches. The MDQR framework demonstrates remarkable computational efficiency, achieving an inference time\footnote{The computational experiments were conducted on an AMD EPYC 7413 server processor operating at 3.16 GHz with 48 threads across 24 cores and 256MB of L3 cache.} of $0.037$ milliseconds per event. This performance translates to just 1.95 minutes for simulating an entire trading day's worth of events\footnote{Reported times account only for event generation and do not include the computational overhead of updating the order book through the matching engine, which is common to all approaches.}, representing a significant advantage over existing approaches. In contrast, while architectures like LOBGAN~\cite{sirignano2019universal} and RNN~\cite{hultin2023generative} may offer sophisticated feature extraction capabilities, they require substantially more computational resources, with daily generation times ranging from 5.5 to 29 minutes.

These performance considerations motivated our choice of a simple MLP architecture over potentially more sophisticated approaches. While advanced architectures might offer enhanced feature extraction capabilities, the specific requirements of reinforcement learning applications necessitate finding an optimal balance between model complexity and computational efficiency. This underscores the importance of considering the intended use case when selecting an appropriate modeling architecture.\\

\subsubsection{Order Sizes Distribution}

A key innovation of the MDQR framework lies in its treatment of order size modeling. Unlike the SAQR model, which relies on independent sampling mechanisms, MDQR implements a conditional approach that considers the broader market context when generating order sizes. This methodology allows the model to capture the subtle interplay between market conditions and traders' size choices, reflecting how market participants adapt their order sizes based on prevailing market conditions and order book state.

\begin{figure}[H]
    \centering
    \includegraphics[width=0.45\textwidth]{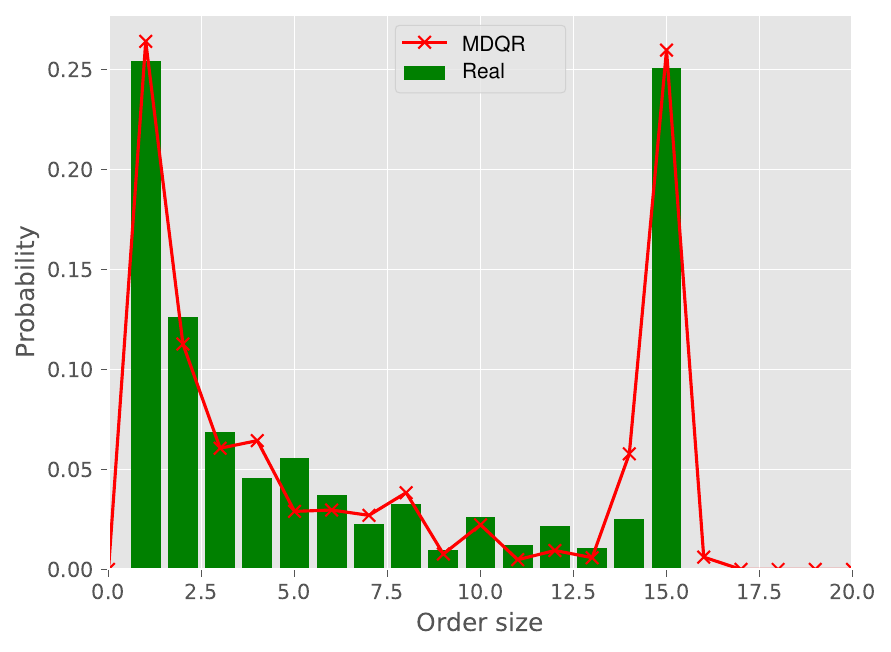}
    \includegraphics[width=0.45\textwidth]{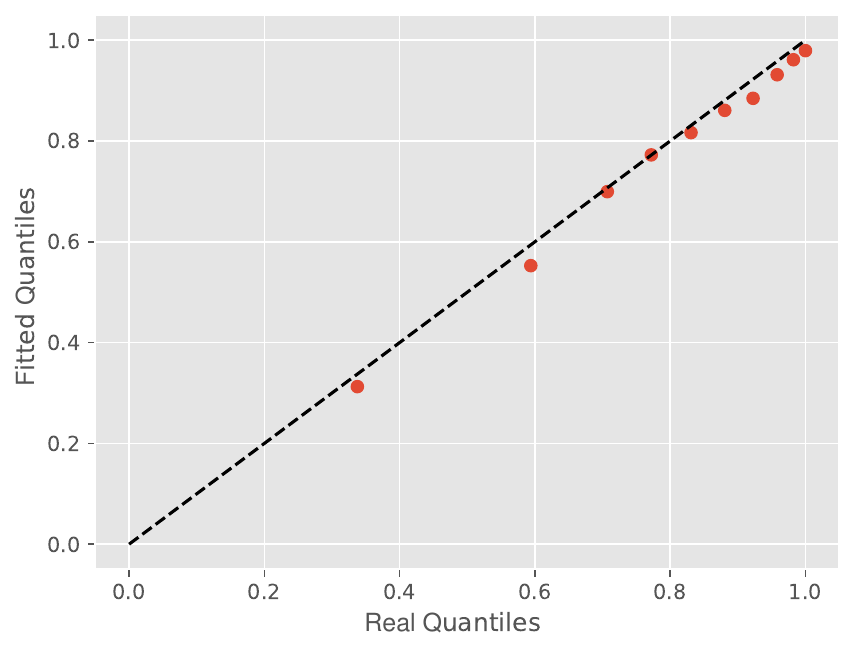}
\caption{Fitted distribution (left) and the corresponding QQ-plot (right) for trade orders arriving when the size of the queue is 15.}
\label{fig:trade_sizes}
\end{figure}

\begin{figure}[H]
    \centering
    \includegraphics[width=0.4\textwidth]{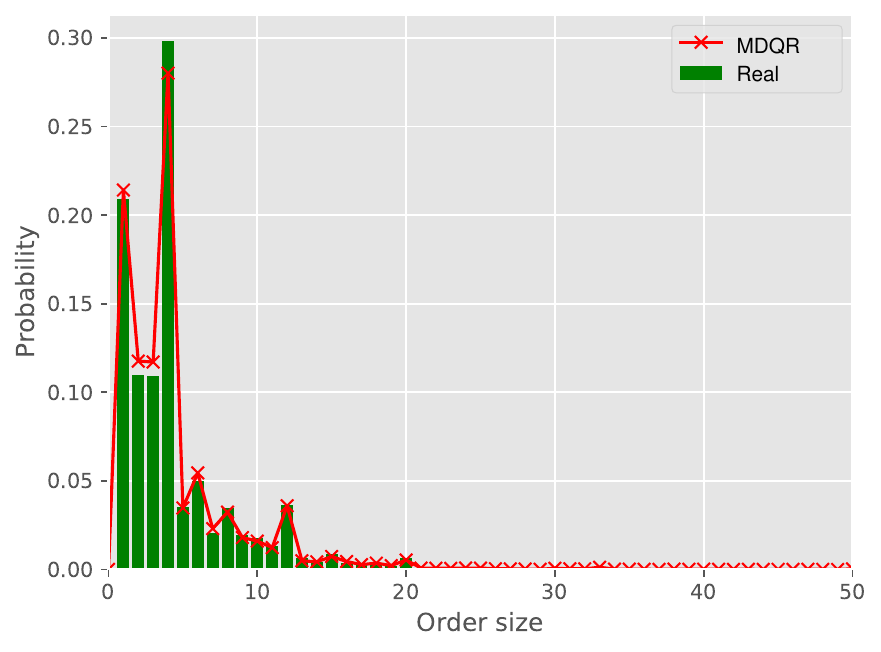}
    \includegraphics[width=0.4\textwidth]{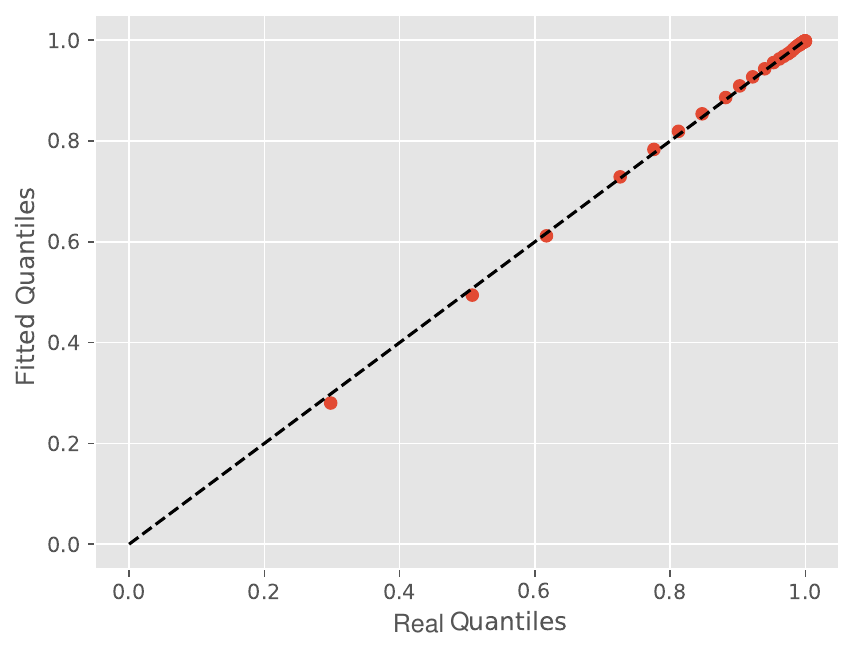}
\caption{Fitted distribution (left) and the corresponding QQ-plot (right) for limit orders arriving when the size of the queue is 20.}
\label{fig:limit_sizes}
\end{figure}




The distribution of order sizes exhibits distinct patterns depending on both the order type and market conditions. Figures~\ref{fig:trade_sizes} and~\ref{fig:limit_sizes} illustrate this complexity through two representative cases. Figure~\ref{fig:trade_sizes} shows the distribution of trade order sizes when the queue size is 15 lots, revealing a specific pattern with notable peaks at certain size values (especially 1 lot and 15 lots). In contrast, Figure~\ref{fig:limit_sizes} presents the distribution of limit order sizes for a queue size of 20 lots, displaying a different structure with higher concentration at smaller sizes. The MDQR sizing model appears to capture these distinct conditional distributions, as evidenced by the close alignment between empirical and simulated patterns in both cases. The corresponding Q-Q plots support this observation, showing consistent agreement across the range of order sizes for both trade and limit orders.

\begin{figure}[H]
    \centering
    \includegraphics[width=0.62\textwidth, trim=0cm 0cm 0cm 1cm, clip]{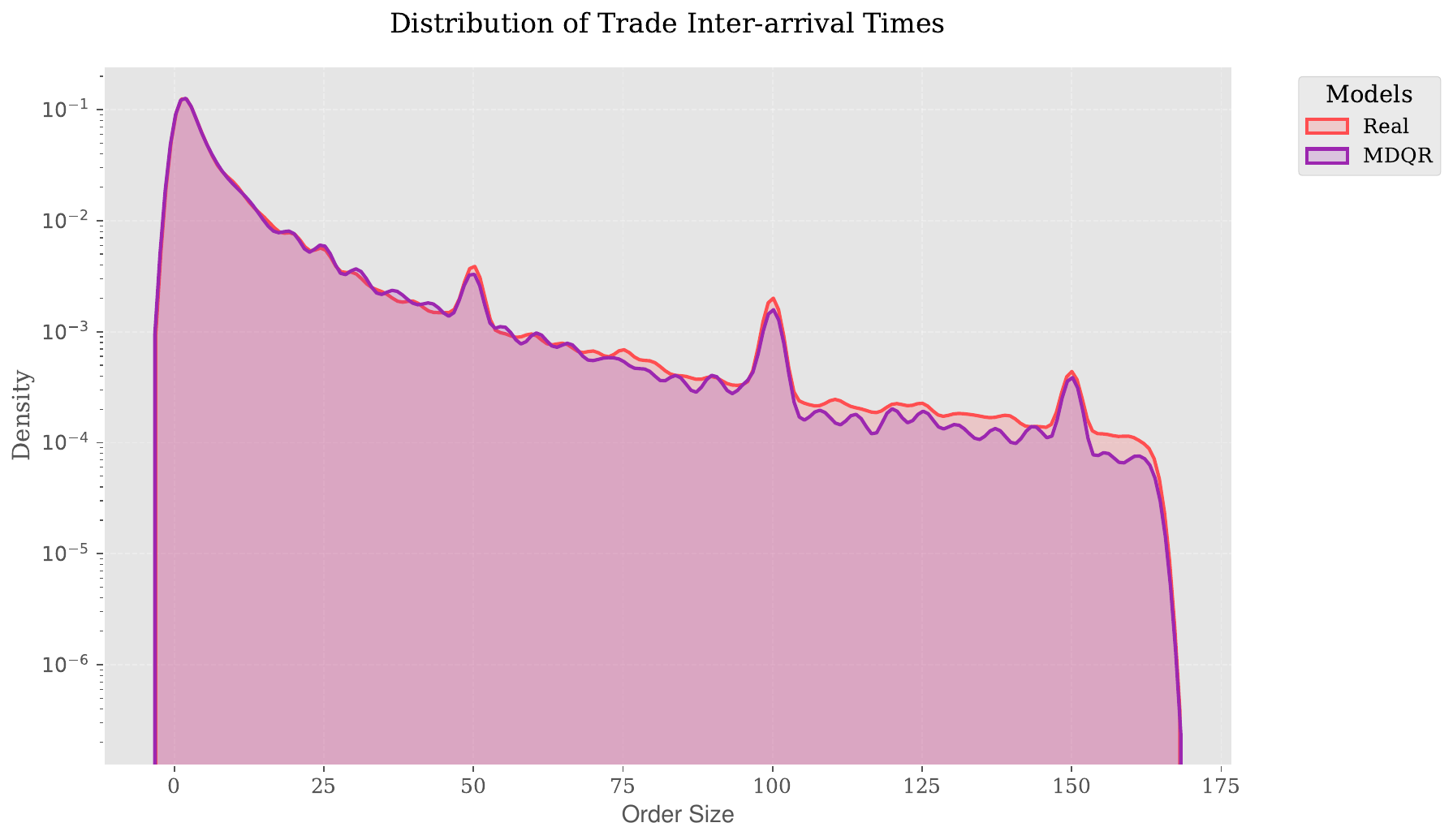}
    \hfill
    \includegraphics[width=0.37\textwidth, trim=0cm 0cm 0cm 1cm, clip]{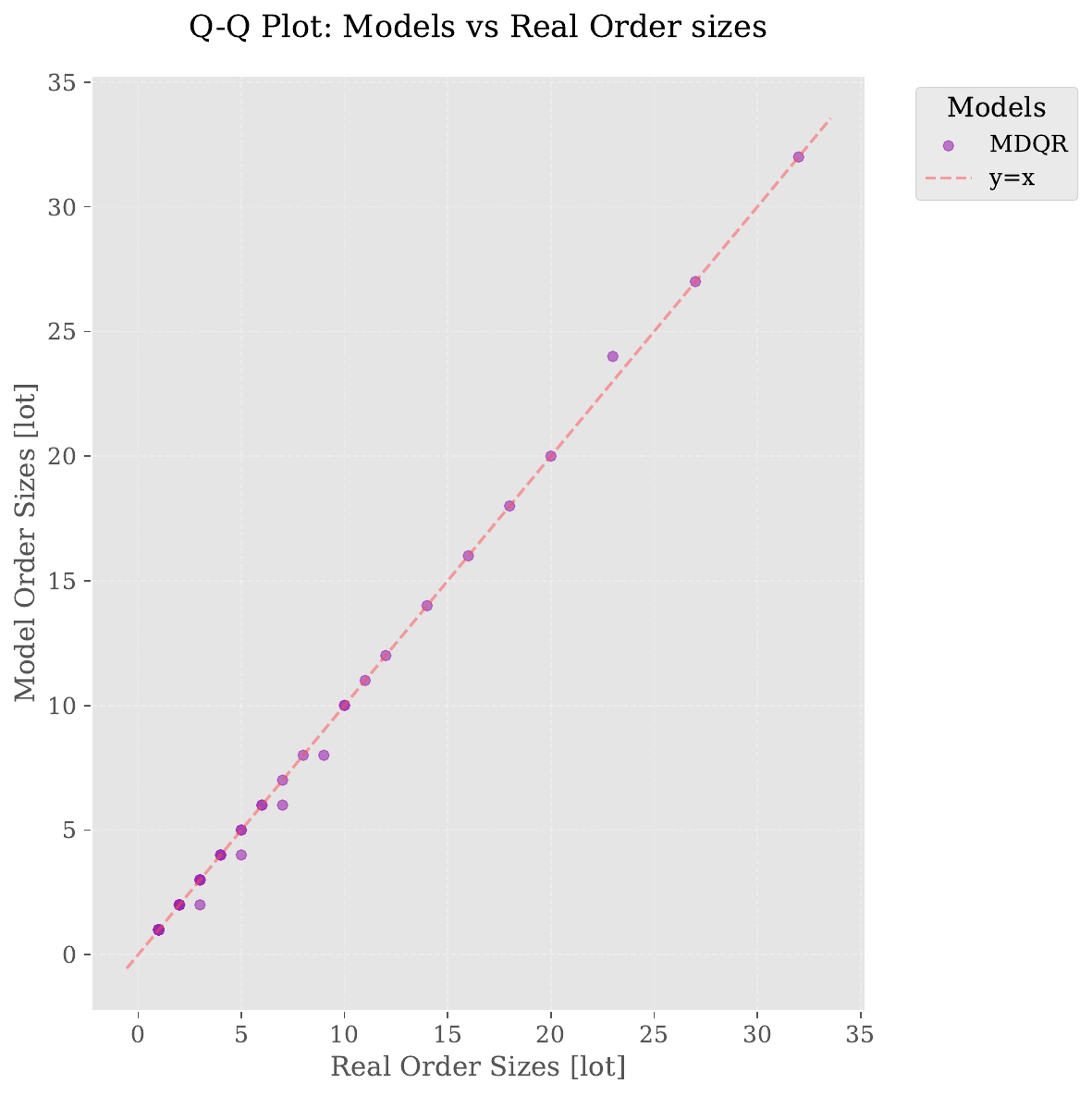}
\caption{\textbf{Left:} Stationary distribution of order sizes for real data and the MDQR model, illustrating the similarity between the two distributions. \textbf{Right:} Q-Q plot comparing the stationary distribution of order sizes generated by the MDQR model with real data, demonstrating the level of agreement between the simulated and actual distributions.}
    \label{fig:sizes_stationary}

\end{figure}

Figure~\ref{fig:sizes_stationary} presents evidence of the model's effectiveness in capturing overall order size dynamics. The left panel displays the stationary distribution of order sizes, where the MDQR-generated distribution mirrors the historical pattern across the entire range of observed sizes. This macroscopic alignment suggests that the model internalizes the factors influencing traders' size decisions. The Q-Q plot in the right panel provides quantitative validation of this distributional accuracy, showing alignment between simulated and empirical quantiles.

These results indicate that the MDQR sizing component can reproduce order size distributions at different scales. The model appears to capture both the specific conditional distributions that arise under particular market conditions and the overall stationary distribution of sizes, suggesting its utility for realistic market simulation.

\subsection{Analysis of Ratio of Limit Orders Filling}

The MDQR simulator enables us to investigate a crucial question faced by market participants: how order fill rate varies with different order characteristics. When seeking to acquire (or sell) a quantity $q$ of an asset, traders can either submit a market order at the best available price, or place a limit order at their desired price level and wait for execution. While limit orders potentially offer better execution prices, they carry the risk of non-execution or adverse price movements.

We study the fill ratio---the rate of partial or complete order execution---as a function of three key characteristics:
\begin{itemize}
    \item Order quantity ($q_{\text{order}}$): the size of the posted order;
    \item Order lifetime ($\tau$): the duration before cancellation\footnote{While traders typically use stop orders, which convert limit orders to market orders after a specified time or price threshold, to manage order lifetime, we focus solely on limit order outcomes within the specified lifetime.};
    \item Price level ($l$): the number of ticks away from the best same-side quote.
\end{itemize}

For our analysis, we consider:
\begin{itemize}
    \item Quantities $q_{\textrm{order}} \in \{1, 2, 5, 10, 20, 50, 60\}$ lots (corresponding approximately to the \\$\{0, 30\%, 50\%, 80\%, 90\%, 95\%, 99\%\}$ percentiles of order sizes of limit orders in the historical data);
    \item Time periods (order lifetimes) $\tau \in \{1, 5, 10, 30, 60, 300\}$ seconds;
    \item Levels $l \in \{0, 1, 2, 3, 4\}$ ticks from the best quote.
\end{itemize}

For each parameter combination, we simulate approximately 2,500 orders. The fill ratio for each order is computed as:
\begin{equation*}
   \textrm{Fill Ratio} = 1 - \frac{q_{\textrm{remaining}}}{q_{\textrm{order}}}
\end{equation*}
This ratio equals 1 when the order is completely filled within the specified period, 0 when no execution occurs, and takes intermediate values for partial executions.

\begin{figure}[H]
    \centering
    \includegraphics[width=\textwidth]{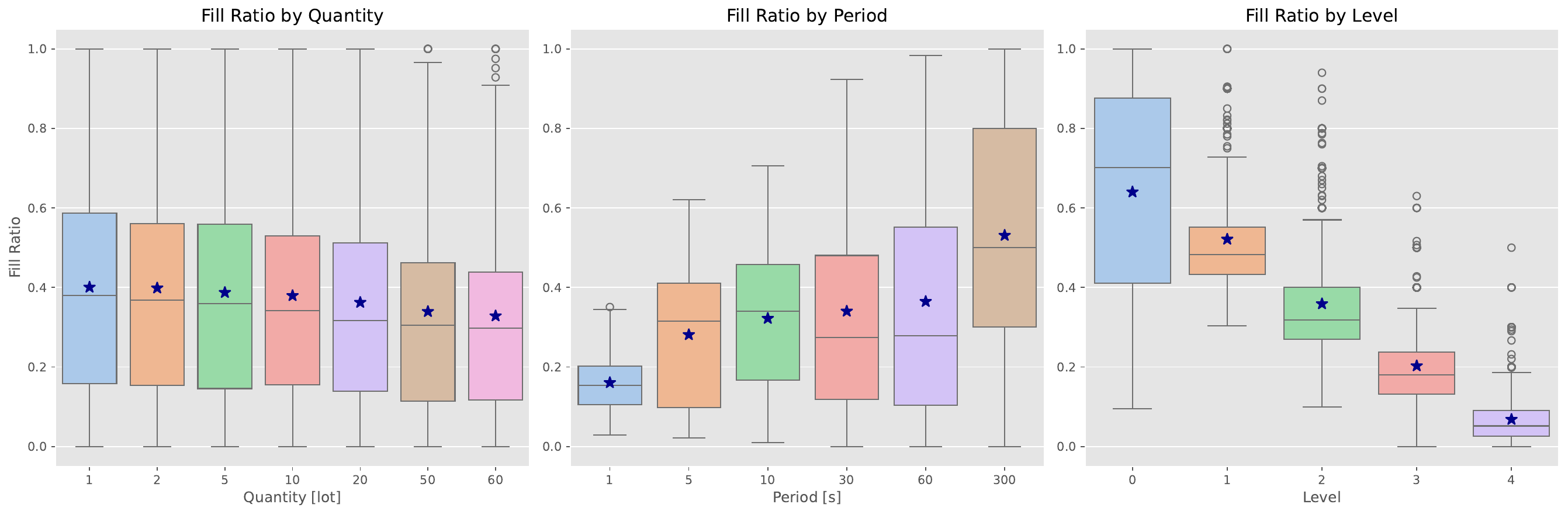}
    \caption{Distribution of fill ratios across different parameter values. Box plots show the quartiles, whiskers extend to the most extreme non-outlier points, and dark blue stars indicate mean values. The quantity plot (left) demonstrates how larger lot sizes affect fill ratio, the period plot (center) shows the relationship between trading frequency and fill rate, and the level plot (right) illustrates the impact of different price levels on fill ratio.}
    \label{fig:fill_prob_dist}
\end{figure}

\begin{figure}[htbp]
    \centering
    \begin{subfigure}[b]{0.32\textwidth}
        \includegraphics[width=\textwidth]{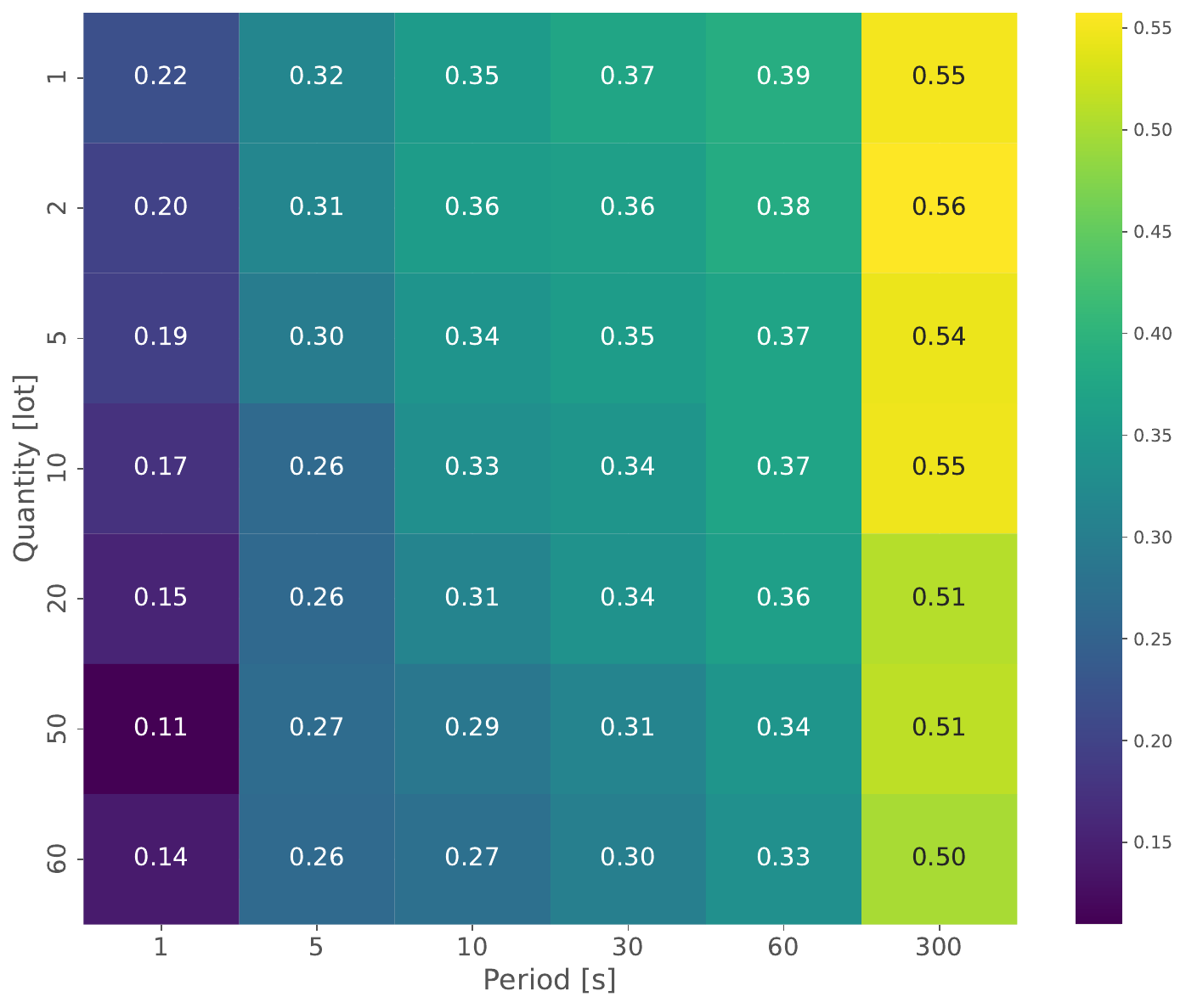}
    \end{subfigure}
    \hfill
    \begin{subfigure}[b]{0.32\textwidth}
        \includegraphics[width=\textwidth]{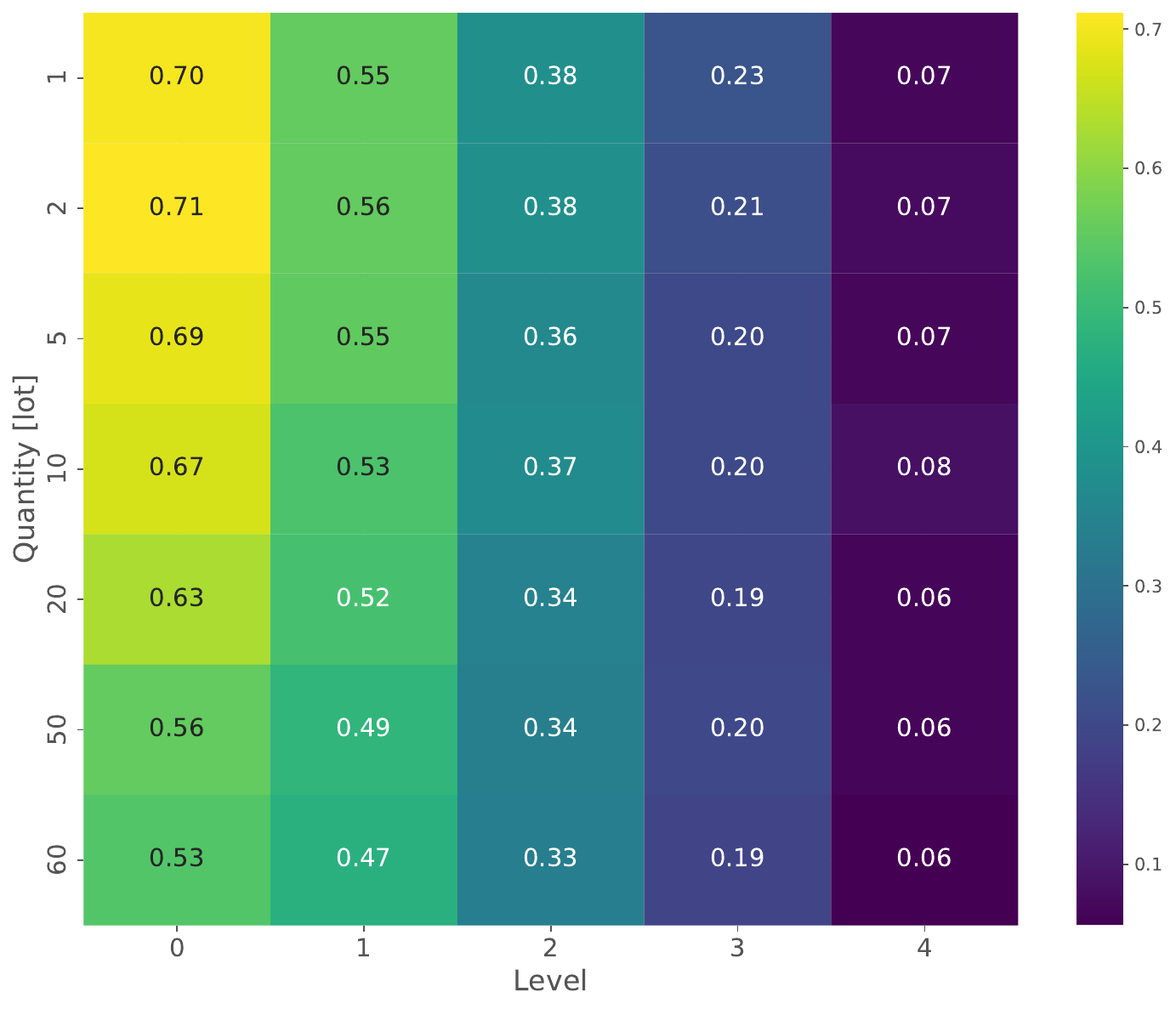}
    \end{subfigure}
    \hfill
    \begin{subfigure}[b]{0.32\textwidth}
        \includegraphics[width=\textwidth]{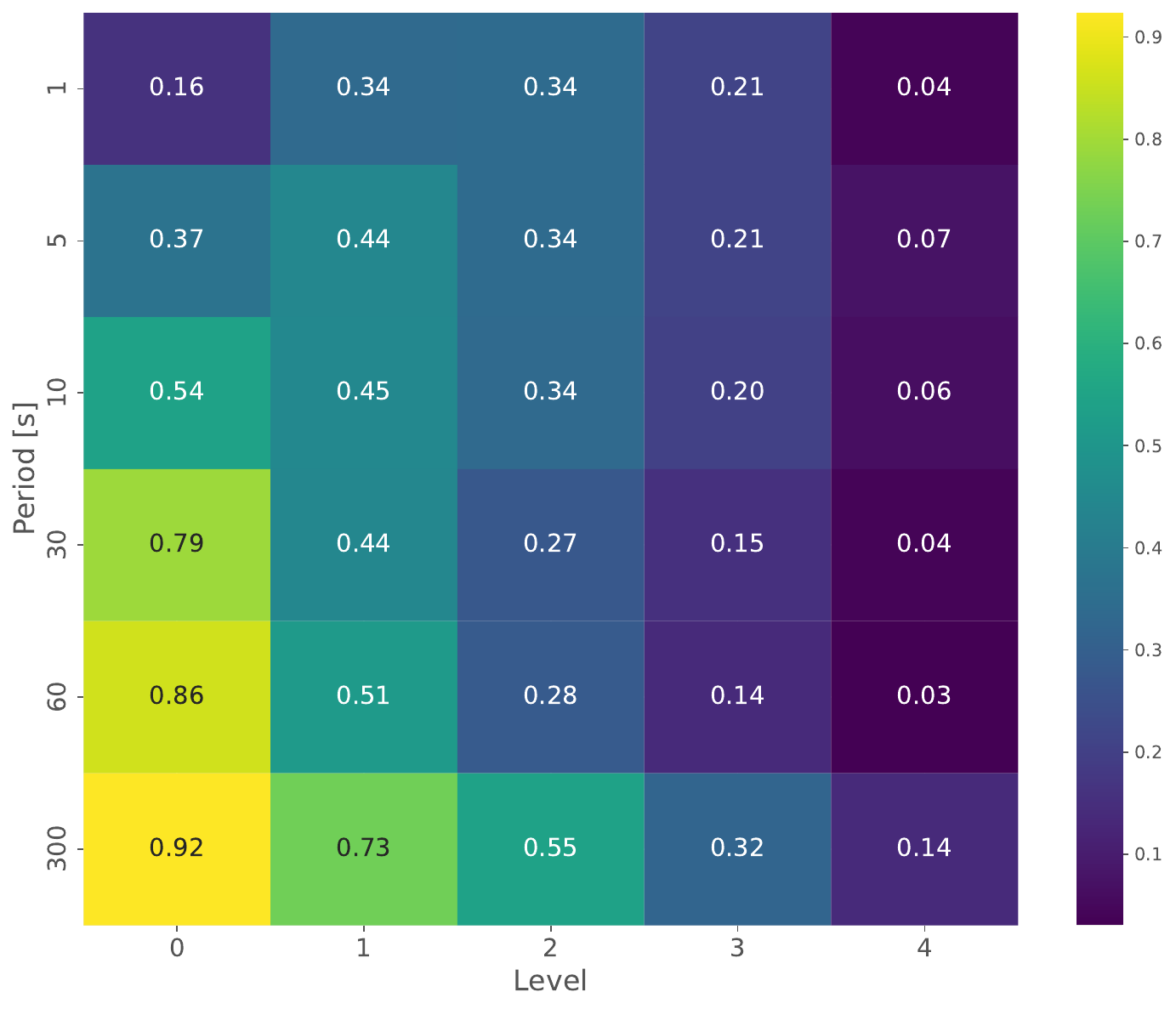}
    \end{subfigure}
    \caption{Heatmaps showing the mean fill ratio for different parameter combinations. Left: Quantity vs Period; Center: Quantity vs Level; Right: Period vs Level. The color scale indicates the fill ratio, with darker red representing higher probabilities and darker blue representing lower probabilities. Values in each cell represent the mean fill ratio for that specific parameter combination.}
    \label{fig:heatmaps_fill_proba}
\end{figure}

Figure~\ref{fig:fill_prob_dist} presents the distribution of fill ratios through box plots for each characteristic, with mean values indicated by dark blue stars. The plots reveal that fill ratio decreases with order size and price level, while increasing with order lifetime. Figure~\ref{fig:heatmaps_fill_proba} further illustrates these relationships through heat maps, showing particularly strong level-dependence and a clear trade-off between quantity and execution probability.

\begin{table}[h]
\centering
\caption{Correlation coefficients between order characteristics and fill ratio}
\label{tab:fill_prob_corr}
\begin{tabular}{lc}
\hline
\textbf{Parameter} & \textbf{Correlation} \\
\hline
Level & -0.79 \\
Period & 0.32 \\
Quantity & -0.10 \\
\hline
\end{tabular}
\end{table}

The correlation coefficients in Table \ref{tab:fill_prob_corr} reveal the relative importance of each characteristic in determining fill ratio. Price level shows the strongest relationship with a substantial negative correlation (-0.79), indicating it is the primary driver of execution likelihood. Order lifetime exhibits a moderate positive correlation (0.32), suggesting longer waiting periods improve execution chances. Order quantity has a minor negative impact (-0.10), showing only a slight decrease in fill ratio as order size increases.

\begin{figure}[H]
    \centering
    \begin{subfigure}[b]{0.45\textwidth}
        \includegraphics[width=\textwidth]{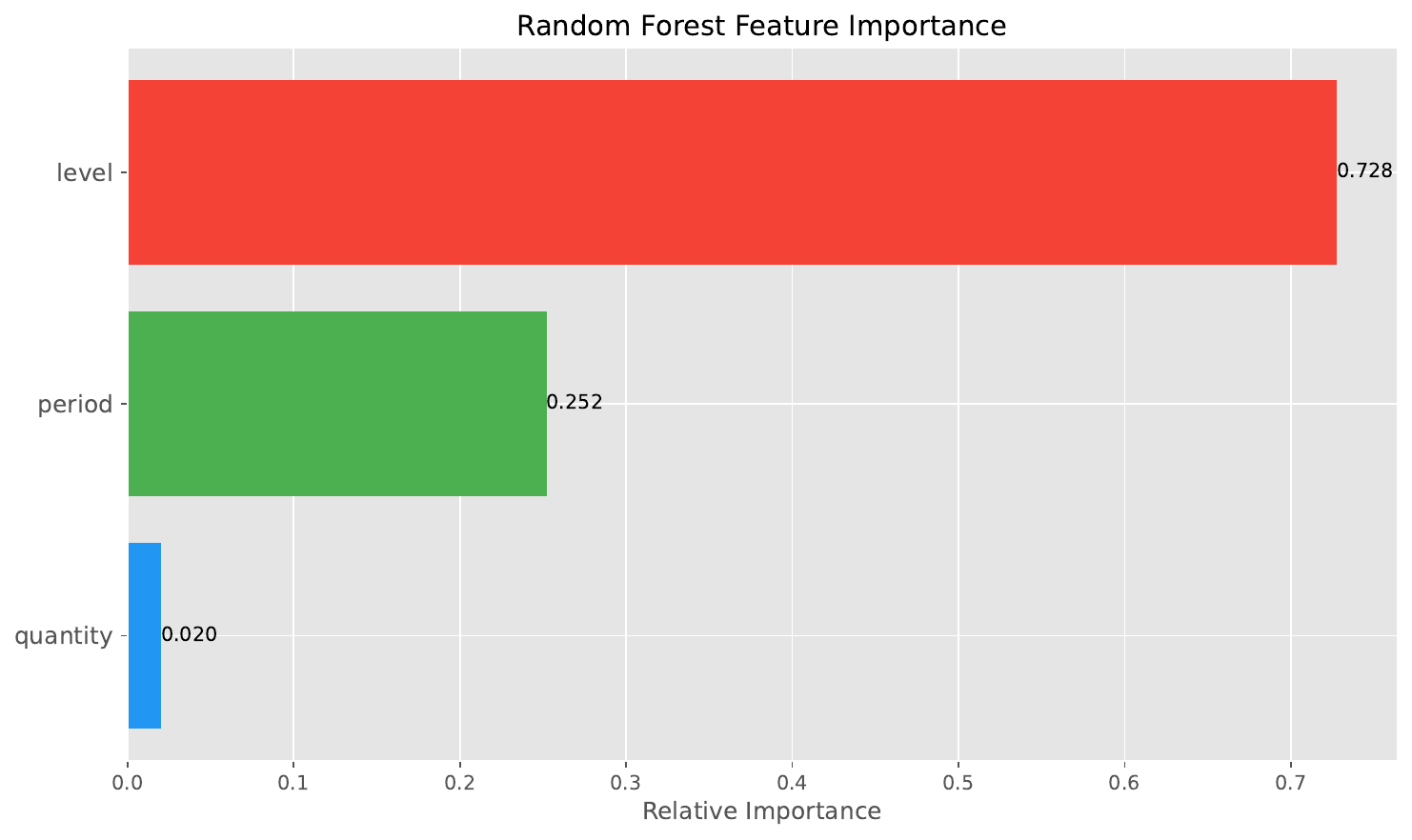}
    \end{subfigure}
    \hfill
    \begin{subfigure}[b]{0.45\textwidth}
        \includegraphics[width=\textwidth]{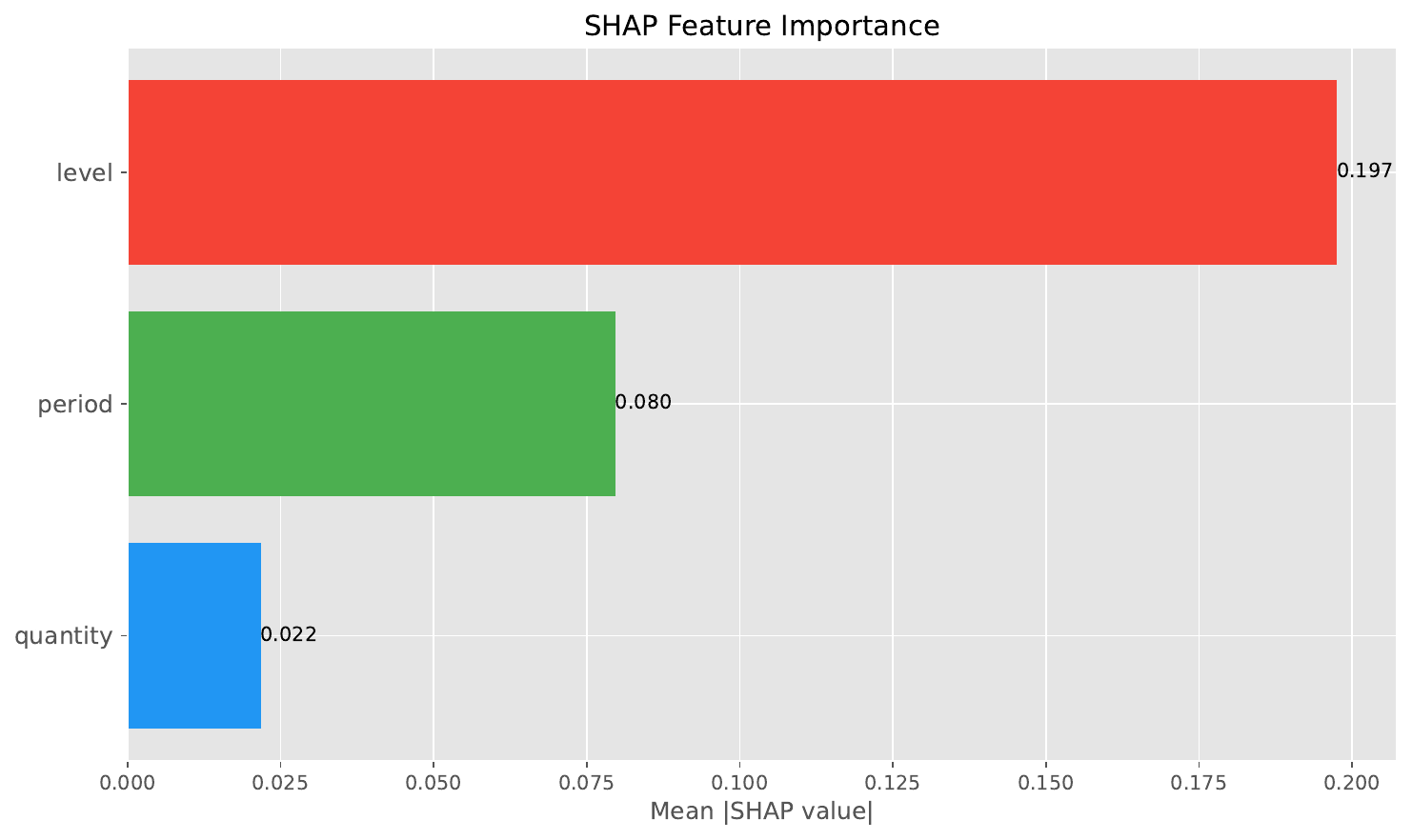}
    \end{subfigure}
    \caption{Feature importance analysis using two different methods. Left: Random Forest feature importance, showing the relative importance of each parameter based on the decrease in model performance when the feature is permuted. Right: SHAP (SHapley Additive exPlanations) values, representing the average magnitude of each feature's contribution to the model predictions.}
    \label{fig:feature_importance}
\end{figure}

While correlation coefficients provide insights into linear relationships between variables, they may not capture more complex, non-linear dependencies. To obtain a more comprehensive understanding of feature importance, we employed two additional analysis techniques: Random Forest feature importance and SHAP (SHapley Additive exPlanations) values. Figure~\ref{fig:feature_importance} presents the results of both approaches, showing remarkable consistency in their findings. Both methods confirm the dominant role of price level in determining fill ratio, followed by a moderate influence of order lifetime, while order quantity exhibits the lowest impact.

\section{Comparison with Recent Deep Learning Approaches}

Our deep learning generalization of the QR model contributes to a growing body of research leveraging neural networks for limit order book simulation. Notable recent approaches include the conditional GAN framework of~\cite{coletta2022conditional}, the Wasserstein GAN with gradient penalty proposed by~\cite{cont2023simulation}, and the multi-model RNN architecture developed by~\cite{hultin2023generative}, which employs separate networks for event type, price level, size, and arrival time generation.

These approaches demonstrate remarkable capability in reproducing market stylized facts. ~\cite{coletta2022conditional} successfully captures the characteristic ``concave-convex'' shape of large order execution impact, a feature also observed in~\cite{hultin2023generative} and our MDQR model. While~\cite{cont2023simulation} does not fully capture the relaxation phase post-execution, it effectively reproduces the square-root law of market impact, demonstrating that maximum impact scales with the square root of executed position size—a property our model also verifies.

In terms of order book structure,~\cite{hultin2023generative} and MDQR both accurately reproduce the average shape across different price levels, while~\cite{cont2023simulation} extends this to capture the complete distribution of queue sizes. Additionally, our MDQR simulator demonstrates strong performance in short-horizon mid-price movement prediction, achieving comparable or superior results to specialized prediction models, aligning with similar findings by~\citet{hultin2023generative}.

While these approaches share common objectives, they offer complementary strengths. The GAN-based models of~\cite{coletta2022conditional} and~\cite{cont2023simulation}, which are architecturally designed for generative tasks, face training stability challenges requiring careful calibration. These models generate order book states at fixed intervals (e.g., 10-second snapshots in~\cite{cont2023simulation}), which proves advantageous when high-frequency precision is unnecessary but limits access to detailed order flow information between snapshots, which limits usage of such information in the modeling.

In contrast,~\cite{hultin2023generative} and MDQR model event-by-event arrival dynamics. The MDQR framework, while not yet validated for small-tick assets, contrasts with~\cite{hultin2023generative} and~\cite{coletta2022conditional}, which demonstrate effectiveness in reproducing spread distributions and regime transitions in such markets. Furthermore,~\cite{coletta2022conditional},~\cite{cont2023simulation}, and MDQR employ manually crafted state spaces, while~\cite{hultin2023generative} leverages automatic feature extraction through LSTM architecture. This distinction presents a trade-off between model complexity control and feature engineering requirements—our implementation, for instance, required several iterations before incorporating trade imbalance features crucial for accurate market impact profiles.

Computational efficiency varies significantly across approaches. MDQR achieves superior inference speed due to its streamlined architecture, while~\cite{hultin2023generative}'s four-model structure results in longer computation times. Our model provides an interpretable framework based on Poisson event arrival while maintaining the ability to reproduce key market stylized facts with efficient simulation capabilities.

The selection of an appropriate model depends on specific use cases. MDQR proves suitable for high-frequency simulation of liquid assets, particularly when computational efficiency is most important.~\cite{cont2023simulation}'s approach may be preferable for lower-frequency applications, despite potential limitations in order flow information. For small-tick assets,~\cite{cont2023simulation} and~\cite{hultin2023generative} demonstrate particular effectiveness. When automatic pattern extraction takes precedence over simulation speed,~\cite{hultin2023generative}'s approach offers distinct advantages through its LSTM architecture.

\section{Conclusion}

This paper presents the Multidimensional Deep Queue-Reactive (MDQR) model, developed through a systematic enhancement of the QR framework. Our initial investigation with the DQR model demonstrated how deep learning techniques could effectively generalize the state space of queue-reactive models, with each additional feature capturing new market properties in the simulated dynamics. Building on these insights, we developed the MDQR framework, which further extends this approach by relaxing the queue independence assumption and introducing a more sophisticated treatment of order sizes.

Our analysis using Bund futures market data suggests the model can capture various market characteristics, including market impact profiles, event-type dependencies across order book sides, and mid-price movement patterns, among others. The MDQR framework incorporates multiple price levels and market state features while retaining computational efficiency. This approach allows for the inclusion of relevant market features in the state space, helping to reproduce observed market behavior in simulations. By maintaining its connection to point processes, the model provides insights into the relationships between market features and order flow patterns. This work contributes to the development of market simulation tools by combining elements from queue-reactive models and deep learning methods. The resulting framework aims to balance model sophistication with practical implementation, while remaining adaptable to evolving market dynamics.

The current implementation shows promising results for liquid and large-tick assets such as the Bund futures and opens several directions for future research. Testing the model's applicability to small-tick assets with different market characteristics would help validate its broader utility. Additionally, the model's computational efficiency makes it particularly suitable for reinforcement learning applications in finance. Future work will focus on leveraging this framework for developing automated trading strategies, particularly in areas such as market making and optimal execution.





\bibliographystyle{elsarticle-num-names}
\bibliography{cas-refs}
 










\end{document}